\documentclass{mn2e}
\usepackage{graphicx}
\usepackage{url}
\usepackage{multirow}
\usepackage{multicol}
\usepackage{caption}
\usepackage{float}
\newcommand{\gtsim}{\mbox{{\raisebox{-0.4ex}{$\stackrel{>}{{\scriptstyle\sim}}  
$}}}}

\title[The galaxy UV luminosity function at $z \simeq 2-4$]{The galaxy UV luminosity function at ${\bf z\simeq 2}$\,-\,4; new results on faint-end slope and the evolution of luminosity density}
\author[Parsa et al.]{Shaghayegh Parsa$^{1}$\thanks{E-mail: shp@roe.ac.uk}, 
James S. Dunlop$^{1}$\thanks{E-mail: jsd@roe.ac.uk}, Ross J. McLure$^{1}$, Alice Mortlock$^{1}$\\
$^{1}$Institute for Astronomy, University of Edinburgh, Royal Observatory, Edinburgh, EH9 3HJ}

\begin{document}


\pagerange{\pageref{firstpage}--\pageref{lastpage}} \pubyear{2014}

\maketitle

\label{firstpage}

\begin{abstract}
We present a new, robust measurement of the evolving rest-frame ultraviolet (UV) 
galaxy luminosity function (LF) 
over the key redshift range from $z \simeq 2$ to $z \simeq 4$. Our results are based on the high dynamic 
range provided by combining the Hubble Ultra Deep Field (HUDF), CANDELS/GOODS-South, and UltraVISTA/COSMOS 
surveys. We utilise the unparalleled multi-frequency photometry available in this survey 
`wedding cake' to compile complete galaxy samples at $z \simeq 2,3,4$ via photometric redshifts 
(calibrated against the latest spectroscopy) rather than colour-colour selection, and to determine
accurate rest-frame UV absolute magnitudes ($M_{1500}$) from spectral energy distribution (SED) fitting.
Our new determinations of the UV LF extend from $M_{1500} \simeq -22$ (AB mag) 
down to $M_{1500}$\,=\,$-$14.5, $-$15.5 and $-$16 at $z\simeq$\,2, 3 and 4 respectively (thus reaching 
$\simeq 3 - 4$ magnitudes fainter than previous blank-field studies at $z \simeq 2 - 3$). At $z \simeq 2 - 3$ we find a much shallower
faint-end slope ($\alpha = -1.32 \pm 0.03$) than reported in some previous studies ($\alpha \simeq -1.7$), 
and demonstrate that this new measurement is robust. By $z \simeq 4$ 
the faint-end slope has steepened slightly, to $\alpha = -1.43 \pm 0.04$, 
and we show that these measurements are consistent with the overall evolutionary 
trend from $z = 0$ to $z = 8$. 
Finally, we find that while characteristic number density ($\phi^*$) drops from $z \simeq 2$ to $z \simeq 4$,
characteristic luminosity ($M^*$) brightens by $\simeq 1$\,mag.. This, 
combined with the new flatter faint-end slopes, has 
the consequence that UV luminosity density (and hence unobscured star-formation density) 
peaks at $z \simeq 2.5 - 3$, when the Universe was $\simeq 2.5$\,Gyr old.
\end{abstract}

\begin{keywords}
galaxies: redshifts - galaxies: spectroscopy - galaxies: 
photometry - galaxies: evolution - galaxies: luminosity function
\end{keywords}

\section{Introduction}

Ultraviolet (UV) continuum emission provides the most direct tracer of star-formation activity in a galaxy
(e.g. Kennicutt \& Evans 2012), albeit it must be corrected for the impact of dust obscuration to 
derive complete star-formation rates. This, coupled with the easy access to the rest-frame ultraviolet 
regime ($\lambda = 1500 - 1700$\,\AA) provided by optical observations of galaxies at redshifts $z$ $\gtsim$ $1.5$,
has meant that the determination of the evolving UV galaxy luminosity function (LF) has become a key probe
of galaxy evolution and overall cosmic star-formation history (e.g. Bouwens et al. 2007;
Reddy \& Steidel 2009; Robertson et al. 2010; McLure et al. 2013; Madau \& Dickinson 2014).

All evidence to date, as provided by a wide range of studies discussed later in this paper,
indicates that, at redshifts $z \simeq 2 - 6$, the UV LF is well described by a Schechter function 
(Schechter 1976). However, as with all LF studies, the challenge is to assemble galaxy samples
of adequate size, spanning a large enough dynamic range, and with sufficiently 
accurate/complete redshift information to robustly determine both the bright and faint end of the LF. 
Thus, ideally, large-area surveys are required to adequately sample the bright end of the LF (and mitigate 
the impact of cosmic variance) while very deep, small-area surveys are necessary to yield the data required 
to properly constrain the faint-end slope, $\alpha$. This latter quantity is of particular importance if it is
hoped to perform a reliable luminosity-weighted integral of the LF down to faint magnitude limits to determine 
UV luminosity density ($\rho_{UV}$). 
In practice, galaxy selection is also a key issue, and it is particularly important to understand 
sample completeness as a function of magnitude and redshift, 
especially if simple colour-colour selection techniques are
utilized to select samples of UV luminous star-forming galaxies. 

After the early pioneering studies of UV luminosity density indicated it was rising with lookback 
time out to at least $z \simeq 1.5$ (e.g. Lilly et al. 1996; Madau et al. 1996; Arnouts et al. 2005), 
the first detailed study of the galaxy UV LF at $z \simeq 2 - 3$ was attempted by Reddy \& Steidel (2009). 
This work was based on colour-selection of galaxies from ground-based data, and as such was best suited 
to determining the bright-end of the LF at $M_{UV} < -18$ (AB mag). Nonetheless, a key result of this paper was 
the derivation of an extremely steep-faint slope for the UV LF at these intermediate redshifts, with 
$\alpha = -1.73 \pm 0.07$. 
This result appears to have been confirmed by subsequent, deeper, {\it Hubble Space Telescope} ({\it HST})
studies following the installation of WFC3 in 2009; first Oesch et al. (2010) reported $\alpha = -1.60 \pm 0.21$ 
at $z \simeq 1.7$, and then, very recently, Alavi et al. (2014) used WFC3/UVIS combined with 
the gravitational lensing boost provided by the cluster A1689 to probe the UV LF at 
$z \simeq 2$ down to (lensing corrected) magnitudes $M_{1500} \simeq -13$, yielding $\alpha = -1.74 \pm 0.08$.
 
However, there has not been universal agreement; for example, Hathi et al. (2010) reported $\alpha = -1.17 
\pm 0.40$ at $z \simeq 2.1$, and Sawicki (2012) found $\alpha = -1.47 \pm 0.24$ at $z \simeq 2$. 
In addition, Weisz, Johnson \& Conroy (2014) have recently used 
galactic archaeology techniques to `reverse engineer' the stellar populations found in 
the present-day Local Group galaxy population, and conclude in 
favour of $\alpha = -1.35 \pm 0.12$ at $z = 2 - 3$.

Such results are interesting, but this controversy simply reinforces the importance of undertaking a new, direct 
investigation of the UV LF at this key epoch in cosmic history. Moreover, a new measurement of the 
UV LF at $z = 2 - 4$ is timely, given the huge recent improvement in the necessary multi-frequency imaging 
(and supporting spectroscopy) in key deep HST {\it and} ground-based survey fields.

In this study we have exploited the combined power of the latest optical--infrared data in the 
Hubble Ultra Deep Field (HUDF) (reaching $\simeq 29.5$ mag over 4.5\,arcmin$^2$), the CANDELS/GOODS-S field
(reaching $\simeq 27.5$\,mag over 170\,arcmin$^2$), and the UltraVISTA/COSMOS field (reaching $\simeq 26$\,mag over 
$\simeq 0.5$\,deg$^2$). In recent years, driven by the rapid improvements in the near-infrared depth
arising from the advent of WFC3/IR and VISTA imaging (e.g. Grogin et al. 2011; McCracken et al. 2012; Koekemoer et al. 2013) 
the unparalleled data in these key fields have been very actively exploited in the 
study of the UV LF at higher redshifts, $z = 5 - 8$ (e.g. McLure et al. 2010, 2013; 
Oesch et al. 2010, 2013; Bouwens et al. 2011, 2015; Finkelstein et al. 2010, 2015; Dunlop 2013; Bowler et al. 2014, 2015). 
However, the full multi-frequency 
datasets have not recently been properly applied to revisit the measurement of the UV LF in the redshift range $z = 2 - 4$.
Our new study thus aims to rectify this situation, and to also take advantage of new optical and near-infrared 
spectroscopy (including WFC3/IR grism spectroscopy; Skelton et al. 2014; Morris et al. 2015) to help produce 
the most reliable photometric redshifts (crucial both for robust galaxy sample selection, 
and for the accurate determination of $M_{1500}$ or $M_{1700}$ for each source).

The remainder of this paper is structured as follows. In Section 2 we summarize the available imaging/photometric 
data for our 3-tier `wedding-cake' survey. Next, in Section 3, we summarize the available spectrosocopic data, 
and explain how we have used robust photometric redshift estimation to assemble a combined sample of $\simeq 35,000$ galaxies 
in the redshift range of interest, $1.5 < z < 4.5$.
We then proceed, in Section 4, to analyse the resulting galaxy dataset to derive the UV LF, focussing first on the detailed 
shape (especially the faint-end slope, $\alpha$) of the LF at $z \simeq 2$, and then expanding the analysis up to $z \simeq 4$.
In Section 5 we compare our results to those of other authors, place our findings in the context of published results 
on the evolution of the UV LF up to $z \simeq 8$, and discuss the implications of our derived UV luminosity 
density, $\rho_{UV}$, over the redshift range $z = 2 - 4$. Our conclusions are summarized in Section 6. 

Throughout we assume a flat cosmology with $\Omega_{0}=0.3$, $\Omega_{\Lambda}=0.7$ and $H_{0}=70$\,kms$^{-1}$Mpc$^{-1}$, 
and give all magnitudes in the AB system (Oke 1974; Oke \& Gunn 1983).

\section{Imaging Data and Photometry}

In this section we summarize the properties of the three surveys used in this study, and explain how the 
galaxies were selected, and their multi-frequency photometry measured. 

Throughout we refer to the $HST$ ACS and WFC3/IR filters F435W, F606W, F775W, F814W, F850LP, F098M, 
F105W, F125W, F140W and F160W as $B_{435}$, $V_{606}$, $i_{775}$, $i_{814}$, $z_{850}$, $Y_{098}$, $Y_{105}$, 
$J_{125}$, $J_{140}$ and $H_{160}$ respectively, the VLT VIMOS-U and Hawk-I/K-{215} filters 
as $U$ and $K_{s}$ respectively, the CFHT MegaCam optical filters as $u$, $g$, $r$, $i'$, $z'$, 
the reddest Subaru Suprime-Cam filter as $z$, the four broad-band VISTA near-infrared filters as 
$Y$, $J$, $H$ and $K_s$, and the $Spitzer$ IRAC first two channels as $IRAC_{3.6{\mu}m}$ and $IRAC_{4.5{\mu}m}$. 

The photometric depths of the imaging data in each filter, for each field, 
are summarized for convenience in Table 1, with further details for
each field given in the following subsections.
 
\subsection{HUDF}

The deepest dataset we analyse, which is crucial for constraining the faint end of the galaxy 
UV LF, is the multi-band imaging of the Hubble Ultra Deep Field completed in 2012 
(HUDF12), covering an area of $\simeq 4.5$\,arcmin$^{2}$. This latest comprehensive dataset 
consists of the deepest near-infrared imaging obtained with $HST$ WFC3/IR from the HUDF09 and HUDF12 
programmes (Bouwens et al. 2010; Ellis et al. 2013; Koekemoer et al. 2013; Dunlop et al. 2013; Robertson et al. 2013), 
and the original optical $HST$ ACS imaging (Beckwith et al. 2006) 
supplemented by new deep $i_{814}$ imaging. To maximise wavelength coverage, we have supplemented 
the $HST$ data with the public VLT GOODS-S $U$ imaging (Nonino et al. 2009), 
the Hawk-I $K_s$ imaging from the HUGS programme (Fontana et al. 2014), 
and the deepest available IRAC imaging (McLure et al. 2011; Ashby et al. 2013).

Galaxy detection and photometry from this deep $HST$ imaging dataset was undertaken using  
\textsc{sextractor} v2.8.6 (Bertin \& Arnouts 1996) in dual-image mode with $H_{160}$ as the 
detection image and the \textsc{flux-iso} as the observed isophotal flux. In order to 
obtain consistent resolution-matched photometry, the lower resolution 
$U$, $K_{s}$ and IRAC images were deconfused using the technique described in McLure et al.
(2011).

Our final photometric catalogue for the HUDF includes the photometry of 
2864 sources with $H_{160} < 29.5$, measured in the 
$U$, $B_{435}$, $V_{606}$, $i_{775}$, $i_{814}$, $z_{850}$, $Y_{105}$, $J_{125}$, 
$J_{140}$, $H_{160}$, $K_{s}$, $IRAC_{3.6{\mu}m}$ 
and $IRAC_{4.5{\mu}m}$ bands.

\begin{table}
\centering
\caption{A summary of the photometry used in this study, giving the 5-$\sigma$ detection limits 
in each filter/field as appropriate. The filter names are as summarized at the beginning 
of Section 2. For the $HST$ photometry the depths given refer to 
total magnitudes, as derived from small-aperture magnitudes assuming point-source corrections (see McLure et al. 2013). 
The $Spitzer$ IRAC depths also refer to total magnitudes, as derived using {\sc TPHOT} (Merlin et al. 2015).
The depths for the ground-based photometry are based on 2-arcsec diameter aperture measurements.}

\begin{tabular}{|l|c|c|c|}

\hline
  \multicolumn{1}{|l|}{\textbf{Filter}} & \multicolumn{3}{c|}{\textbf{Survey}}\\ 
\hline
\hline

   & HUDF & CANDELS/ & UltraVISTA/\\
   &      & GOODS-S & COSMOS\\  
 \hline 

$U$ & 28.0 & 28.0 &\\
$u$  & & & 27.0 \\
$B_{435}$ & 29.7 & 28.0 & \\
$g$  & & & 27.1 \\
$V_{606}$ & 30.2 & 28.4 & \\
$r$  & & & 26.6 \\
$i_{775}$ & 29.9 & 27.8 & \\
$i'$ & & & 26.3 \\
$i_{814}$ & 29.8 & & \\
$z'$ & & & 25.4 \\
$z$ & & & 26.4 \\
$z_{850}$ & 29.1 & 27.5 & \\
$Y_{105}$ & 29.7 & 27.9 & \\
$Y$ & & & 25.1\\
$J_{125}$ & 29.2 & 27.7 &  \\
$J$ & & & 24.9\\
$J_{140}$ & 29.2 & & \\
$H_{160}$ & 29.2 & 27.3 & \\
$H$ & & & 24.6\\
$K_{s}$ & 26.5 & 26.5 & 24.8\\
$IRAC_{3.6}$ & 26.5 & 26.5 & 25.2 \\
$IRAC_{4.8}$ & 26.3 & 26.3 & 25.2\\
\hline\end{tabular}
\end{table}

\subsection{CANDELS/GOODS-S}

To provide the next tier of the survey `wedding cake', we have used the publicly-available 
{\it HST} WFC3/IR and {\it HST} ACS imaging 
of the Great Observatories Origins Deep Survey South (GOODS-S) field provided by 
the Cosmic Assembly Near-Infrared Deep Extragalactic Legacy Survey (CANDELS) (Grogin et al. 2011; Koekemoer et al. 2011;
Windhorst et al. 2011),
and the associated pre-existing 
{\it HST} optical (Giavalisco et al. 2004; Riess et al. 2007), ground-based VLT $U$-band (Nonino et al. 
2009) and $K_s$-band (Retzlaff et al. 2010; Fontana et al. 2014), and {\it Spitzer} IRAC
imaging (Ashby et al. 2013), as summarized by Guo et al. (2013). 

Consistent with the production of the HUDF catalogue, the sources were detected, and their isophotal fluxes in the $HST$ bands 
measured using \textsc{sextractor} v2.8.6 (Bertin \& Arnouts 1996) in dual-image mode, again with $H_{160}$ as the detection image.
As described in Guo et al. (2013), in this field the Template FITting (\textsc{tfit}) method
(Laidler et al. 2007) has been applied to generate the matched photometry from the lower angular resolution 
$U$, $K_{s}$ and IRAC imaging. The GOODS-S catalogue provided by Guo et al. (2013) contains 
34930 sources in an area of 173\,arcmin$^{2}$, with photometry 
in the $U$, $B_{435}$, $V_{606}$, $i_{775}$, $z_{850}$, $Y_{098}$, $Y_{105}$, $J_{125}$, 
$H_{160}$, $K_{s}$, $IRAC_{3.6{\mu}m}$ and $IRAC_{4.5{\mu}m}$ bands.

\subsection{UltraVISTA/COSMOS}
Data Release 2 (DR2) of the UltraVISTA survey (McCracken et al. 2012) provides deep near-infrared imaging in 
4 deep strips which overlap $\simeq 0.7$\,deg$^2$ of the area also covered by the {\it HST} imaging of the 
Cosmological Evolution Survey (COSMOS; Scoville et al. 2007), and by 
the Canada-France-Hawaii Telescope Legacy Survey (CFHTLS) MegaCam deep optical imaging. 
As discussed in Bowler et al. (2012, 2014, 2015), for the central square degree covered by the 
CFHTLS D2 imaging, very deep Subaru Suprime-Cam $z$-band imaging has also been obtained (Furusawa et al., in preparation).

The photometry of the sources in the UltraVISTA+COSMOS imaging has been measured in 2$^{\prime \prime}$ diameter apertures using \textsc{sextractor} in dual image mode. 
The detection image used in this case is the $i'-$band image from the T0007 release of the CFHTLS. These fluxes have been converted to 
total using the $i'-$band FLUX\_AUTO parameter, and in this case the new \textsc{tphot} code (Merlin et al. 2015) has been used to obtain 
deconfused, resolution-matched IRAC fluxes.

The final UltraVISTA+COSMOS catalogue utilised here contains 89614 galaxies with $i' < 26$, selected 
from an area of 0.482\,deg$^2$ (reduced to an effective survey area of 0.292\,deg$^2$ after masking 
for bright objects, and diffraction spikes etc), and provides
photometry in the $u$, $g$, $r$, $i'$, and $z'-$bands from 
the CFHTLS, the $z-$band from Suprime-Cam on Subaru, the $Y$, $J$, $H$, and $K_s$ bands from UltraVISTA, and 
deconfused $IRAC_{3.6{\mu}m}$ and $IRAC_{4.5{\mu}m}$ photometry from a combination of the {\it Spitzer} Extended 
Deep Survey (SEDS; PI: Fazio; Ashby et al. 2013) and the \textit{Spitzer} Large Area Survey with Hyper-SuprimeCam 
(SPLASH; PI: Capak).

\begin{figure} 
\centering
\includegraphics[width=0.47\textwidth]{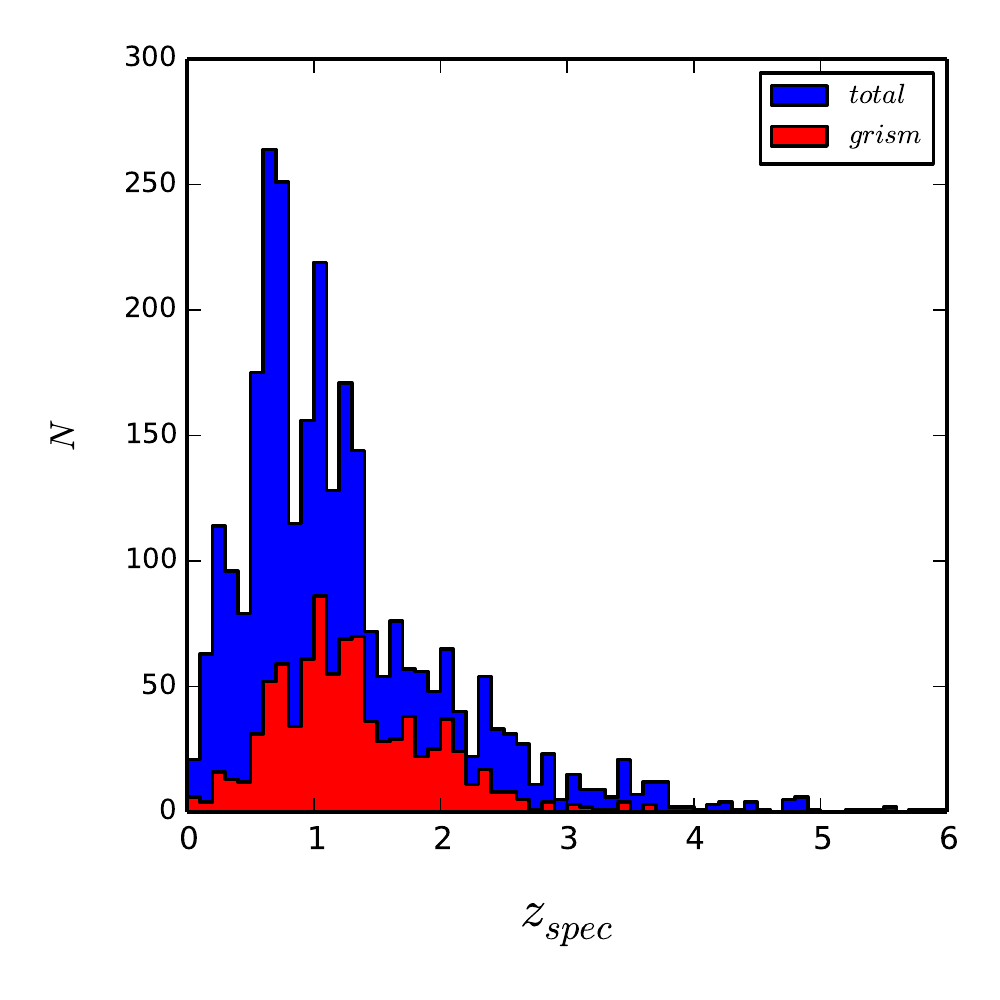}
\caption{The distribution of the 2799 high-quality spectroscopic 
redshifts in the latest spectroscopic sample we have assembled for the 
CANDELS/GOODS-S field. The red histogram indicates the new WFC3/IR near-infrared grism redshifts 
as determined by the 3D-HST survey (Skelton et al. 2014) and CANDELS (Morris et al. 2015), which have helped significantly to fill in the 
`redshift desert'. Of the 2799 objects with high-quality spectroscopic 
redshifts shown here, 218 lie within the HUDF.}
\end{figure}

\section{Redshift Information}

In an ideal world, the galaxy LFs would be derived from large, complete, galaxy samples with complete spectroscopic 
redshift information. However, at the depths of interest here, even semi-complete spectroscopic redshift 
information is clearly (currently) not possible. Consequently, the 
robustness of our final results depends crucially on the reliability and accuracy of photometric redshifts 
derived for the galaxies uncovered for each of the three survey fields discussed above.
  
To test/optimize the derived photometric redshifts, subsets of objects with reliable spectroscopic redshifts are 
required. We have therefore assembled the latest catalogues of robust spectroscopic redshifts in the HUDF, GOODS-S, and COSMOS 
fields. The results of this search are summarized in the first sub-section below. 

We then explain the steps taken to optimize photometric redshifts, the final procedure adopted, and quantify the reliability and 
accuracy of our final photometric redshifts via comparison with the spectroscopic database (with further details and 
comparisons with other published photometric redshift catalogues given in Appendix A).

In the third subsection below we describe our final combined sample of galaxies with redshifts in the range $1.5 < z < 4.5$. 
This sample is used in the LF analyses presented in the remainder of this paper.

\subsection{Spectroscopic Redshifts}

Despite the fact that the HUDF, GOODS-S and COSMOS fields have been targeted by 
several of the most dense and deep spectrosopic surveys ever undertaken, typically at most $\simeq 10$\% 
of the galaxies in our photometric samples possess high-quality spectroscopic redshifts.
Nonetheless, spectroscopic redshift information is crucial for refining and quantifying the 
accuracy (e.g. $\sigma$) and reliability (e.g. number of catastrophic outliers) of photometric 
redshifts.

We have therefore assembled catalogues including the very latest spectroscopic redshift information 
in each field. Within the GOODS-S field (including the HUDF) we have assembled a sample of 2799 
galaxies with high-quality redshifts (218 of which lie within the 
area covered by the WFC3/IR imaging of the HUDF). 
We have confined our selection to only the very highest quality 
flags assigned to the redshifts obtained by each study in the literature, and 
the resulting redshift distribution of our final spectroscopic galaxy sub-sample 
(after removal of any stars or AGN) is shown in Fig.\,1. 
As illustrated by the blue histogram in Fig.\,1, the majority of this spectroscopic 
redshift information (1917 redshifts) 
has been obtained from ground-based optical spectroscopy 
(Balestra et al. 2010; Cimatti et al. 2008; Cristiani et al. 2000; Croom et al. 2001; 
Dohetry et al. 2005; Szokoly et al. 2004; Le F\'{e}vre et al. 2004; 
Mignoli et al. 2005; Roche et al. 2006; Silverman et al. 2010; Strolger et al. 2004; Vanzella et al. 2008). 
However, as illustrated by the red histogram in Fig.\,1,  
it can also been seen that recent $HST$ WFC3/IR near-infrared grism spectroscopy (Skelton et al. 2014; Morris et al. 
2015) has now made an important contribution to the redshift coverage in this deep field (982 redshifts), 
in particular helping to fill in the traditional `redshift desert' between $z \simeq 1.2$ and 
$z \simeq 2$, where relatively few strong emission 
lines are accessible in the optical regime. 

Within the COSMOS field we utilized a sample of 1877 high-quality 
redshifts as provided by the public $z$-COSMOS survey (Lilly et al. 2007).

\begin{figure} 
\centering	
\includegraphics[width=0.435\textwidth]{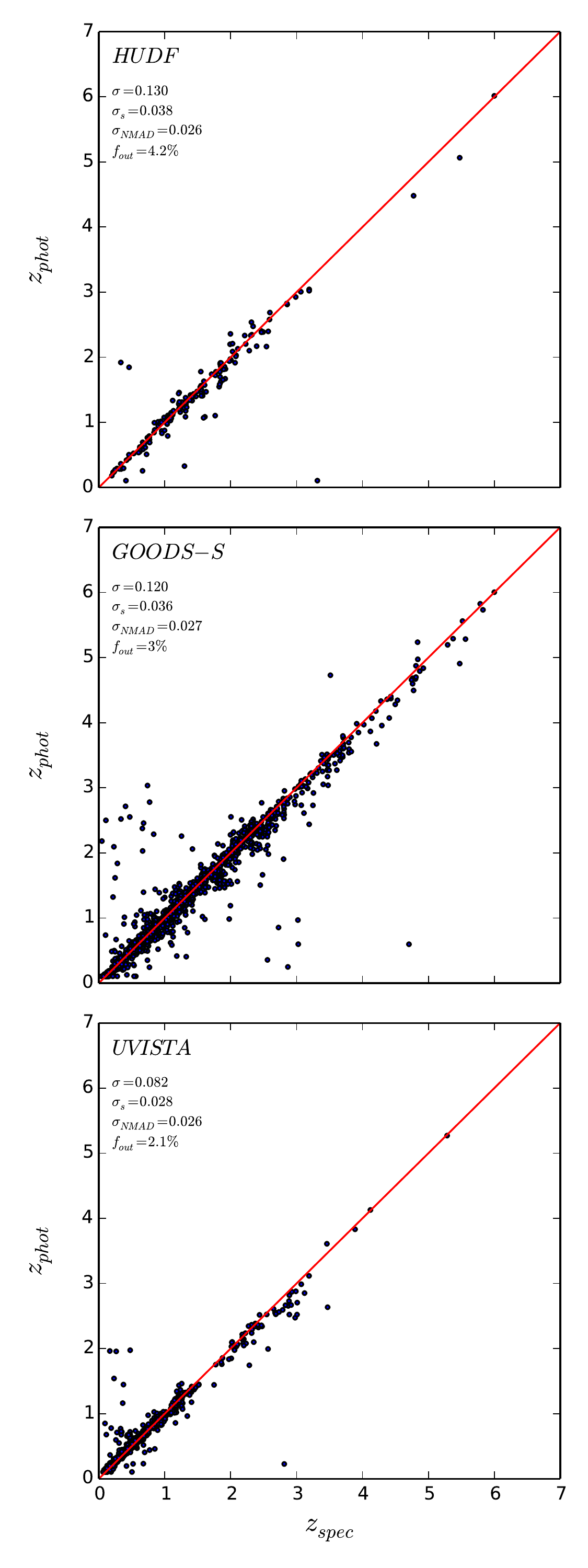}
\caption{The comparison of our new photometric redshifts versus the high-quality spectroscopic redshifts 
in each of the three survey fields. The top plot for the HUDF contains 210 galaxies, the central plot for 
CANDELS/GOODS-S contains 2677 galaxies, while the bottom plot for UltraVISTA/COSMOS contains 1671 galaxies.
As summarized by the statistics in each panel, the accuracy of our photometric redshifts is comparable 
with the very best ever achieved for high-redshift galaxy surveys; see Section 3 and Appendix A for futher details.}
\end{figure}
\begin{figure} 
\centering	
\includegraphics[width=0.435\textwidth]{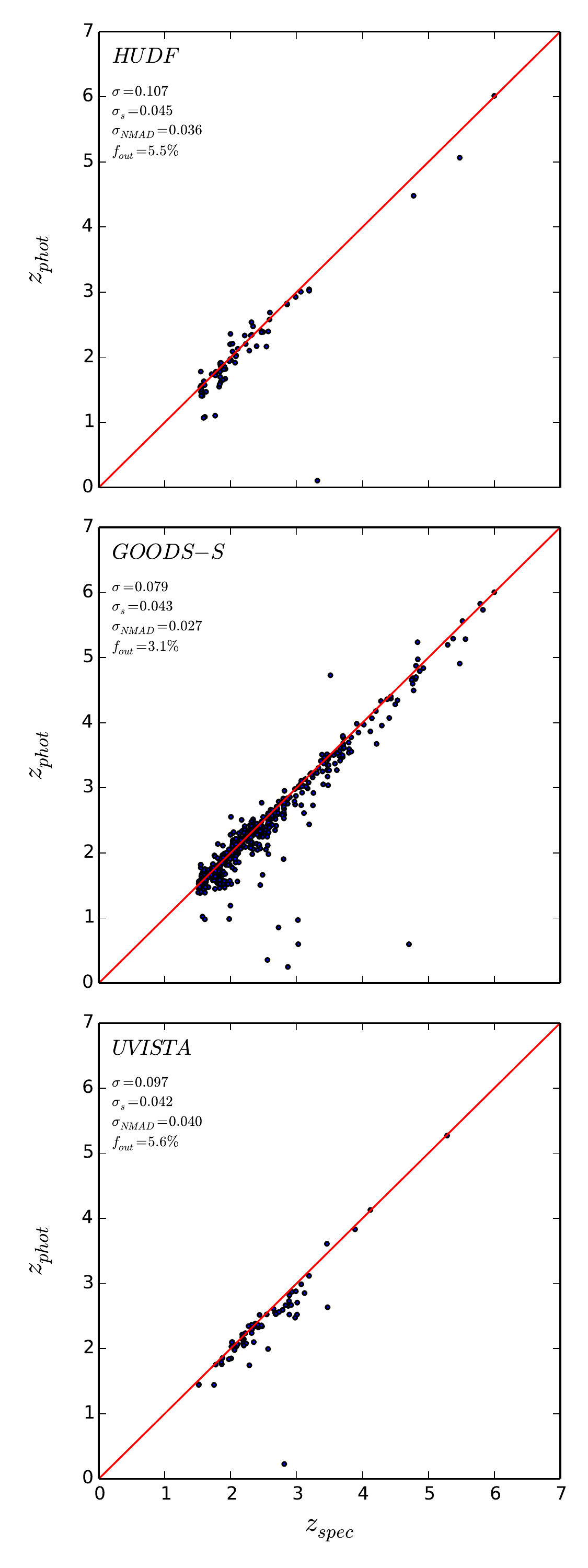}
\caption{A second comparison of our new photometric redshifts versus the high-quality spectroscopic redshifts 
in each of the three survey fields, this time confined to $z > 1.5$. The HUDF plot contains 72 galaxies, the 
CANDELS/GOODS-S plot contains 668 galaxies, while the UltraVISTA/COSMOS plot contains 71 galaxies.
As summarized by the statistics in each panel, the accuracy of our photometric redshifts remains comparable 
with the very best ever achieved for high-redshift galaxy surveys; see Section 3 and Appendix A for futher details.}
\end{figure}

\subsection{Photometric Redshifts}

\subsubsection{Method}

To determine photometric redshifts we used the public galaxy template-fitting 
code {\sc Le Phare}\footnote{\url{http://www.cfht.hawaii.edu/~arnouts/LEPHARE/lephare.html}} 
(PHotometric Analysis for Redshift Estimate; Ilbert et al. 2006). To ensure the proper 
treatment of weak/non-detections, we fitted in flux-density rather than magnitude space. To account 
for dust obscuration/reddening we assumed the dust-attenuation law of Calzetti et al. (2000), 
allowing reddening to vary over the range $0.0 < E(B-V) < 0.5$ in steps of $\Delta{E(B-V)}=0.1$. 
We also included IGM absorption assuming the models of Madau (1995).

For each field we proceeded in four stages. First, to avoid too much weight being placed on individual 
photometric detections, and to allow for remaining low-level systematic errors, we set a minimum of 
error of 3\% on all optical and near-infrared photometry, and a minimum error of 10\% on all 
IRAC photometry. Next, we utilized the galaxy SED templates provided by the evolutionary synthesis 
models of Bruzual \& Charlot (2003) (BC03), without emission lines, and adjusted the photometric zero-points 
until the accuracy of the photometric redshifts was maximised (as judged by comparison with the high-quality 
spectroscopic redshifts discussed above).
After this we explored the use of a range of different galaxy templates, before determining that the 
the PEGASEv2.0 models (Fioc \& Rocca-Volmerange 1999), with emission lines 
switched on, produced the most accurate final photometric redshifts. Finally, we determined the photometric 
redshift for each galaxy by searching the redshift range $z = 0 -10$, and distinguished 
between acceptable and unacceptable photometric redshifts based on an analysis of the distribution of minimum $\chi^2$ 
(resulting in the acceptance of final SED fits with minimum $\chi^2 < 50$ in the fields with 
{\it HST} photometry, and minimum $\chi^2 < 20$ in the UltraVISTA/COSMOS field).
This level of quality control led to the exclusion of $<5$\% of galaxies
in the photometric samples from the final sample with trusted photometric redshifts.  

Below we summarize the results obtained in each field.

\subsubsection{HUDF photometric redshifts}

As outlined above, we refined the HUDF photometric zeropoints by fitting the photometry 
with the BC03 models; the derived 
zero-point offsets were all smaller than 0.1\,mag. After application of {\sc Le Phare} with emission lines, 
we derived acceptable ($\chi^2 < 50$) photometric redshifts for 2730 galaxies with $H < 29.5$ in the HUDF. 

The accuracy of the derived photometric redshifts in the HUDF is illustrated in the top panel 
of Fig.\,2, which shows $z_{phot}$ v $z_{spec}$ for the 210 sources in the field with secure spectroscopic redshifts,
{\it and} acceptable photometric redshifts. The outlier fraction is 4.2\%, and $\sigma_{NMAD} = 0.026$.
For redshifts confined to $z > 1.5$ the $z_{phot}$ v $z_{spec}$ plot is shown in the 
top panel of Fig.\,3 (see Appendix A for further details).  

\subsubsection{CANDELS/GOODS-S photometric redshifts}

To maximise the reliability of the photometric redshifts for this sample, we confined our attention to sources with $H_{160} < 26$ in 
the CANDELS Wide region, and to sources with $H_{160} < 27$ in the CANDELS Deep region (which covers the central 
$\simeq 55$\,arcmin$^2$ of GOODS-S), and then again refined the photometric zeropoints by fitting the photometry 
with the BC03 models; the derived zero-point offsets were again all smaller than 0.1\,mag.

After application of {\sc Le Phare} with emission lines, we derived acceptable ($\chi^2 < 50$) photometric redshifts 
for 10987 galaxies with $H < 27$ in the GOODS-S Deep field, and for 
27460 galaxies with $H < 26$ in the GOODS-S Wide field. 

The accuracy of the derived photometric redshifts in CANDELS/GOODS-S is illustrated in the middle panel 
of Fig.\,2, which shows $z_{phot}$ v $z_{spec}$ for the 2677 sources in the field with secure spectroscopic redshifts,
{\it and} acceptable photometric redshifts. The outlier fraction is 3\%, and $\sigma_{NMAD} = 0.027$.
For redshifts confined to $z > 1.5$ the $z_{phot}$ v $z_{spec}$ plot is shown in the 
middle panel of Fig.\,3 (see Appendix A for further details).

\subsubsection{UltraVISTA/COSMOS photometric redshifts}

As with the $HST$-based catalogues, initially the photometric redshifts 
were computed by fitting the ground-based and {\it Spitzer} photometry to the 
BC03 models, in order to adjust the photometric zero-points 
until maximum redshift accuracy was achieved; all the derived zero-point offsets were again all smaller than 0.1\,mag.

For consistency with the {\it HST} HUDF+GOODS-S data analysis, the 
photometric redshifts and rest-frame absolute UV magnitudes were then again 
recomputed using {\sc Le Phare} with emission lines, yielding acceptable ($\chi^2 < 20$) photometric redshifts 
for 88789 galaxies with $i' < 26$ in the UltraVISTA/COSMOS field.

The accuracy of the derived photometric redshifts in UltraVISTA/COSMOS is illustrated in the bottom panel 
of Fig.\,2, which shows $z_{phot}$ v $z_{spec}$ for the 1671 galaxies in the field with secure spectroscopic redshifts,
{\it and} acceptable photometric redshifts. The outlier fraction is only 2.1\%, and $\sigma_{NMAD} = 0.026$.
For redshifts confined to $z > 1.5$ the $z_{phot}$ v $z_{spec}$ plot is shown in the 
bottom panel of Fig.\,3 (see Appendix A for further details).  

\begin{figure} 
\centering	
\includegraphics[width=0.47\textwidth]{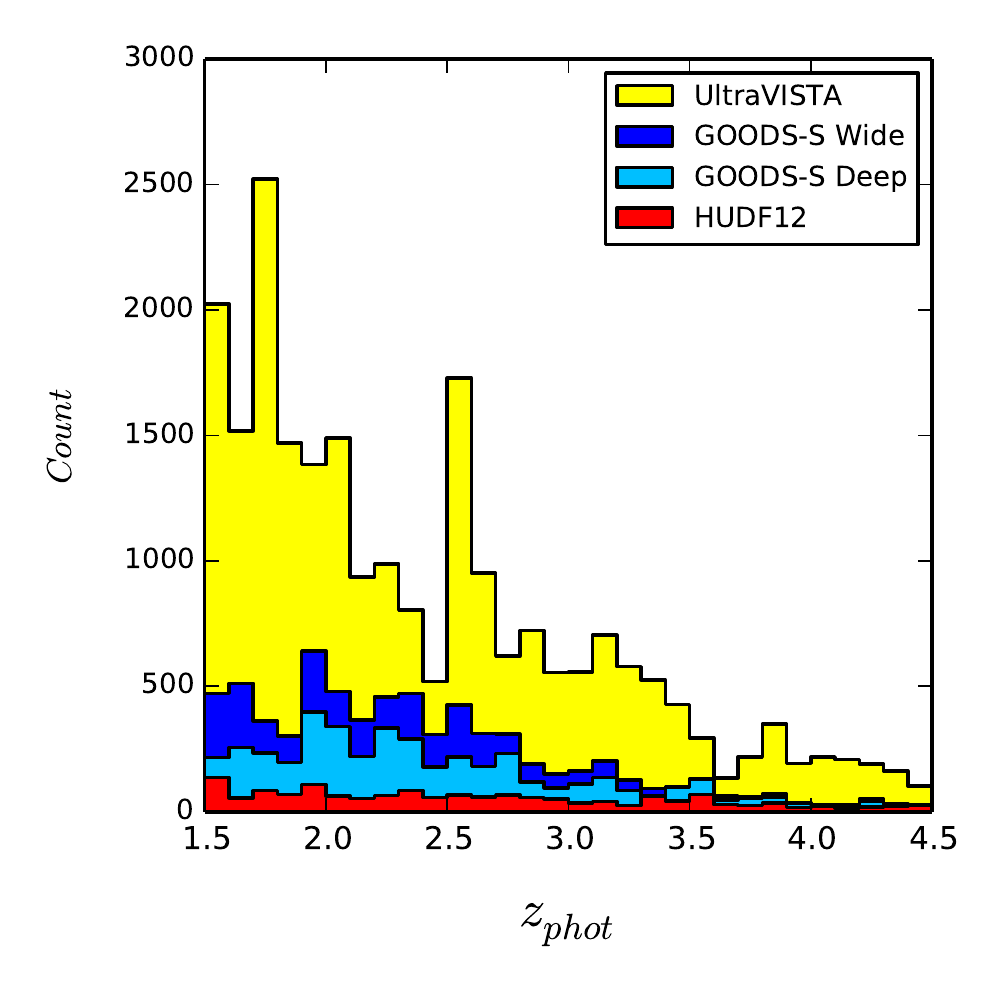}
\caption{The distribution of photometric redshifts for our final galaxy sample comprising a total 
of 36051 galaxies in the redshift range $1.5<z_{phot}<4.5$. The different coloured histograms show the 
redshift distribution subdivided by survey-field/depth, with 1549 galaxies from the HUDF with $H_{160} < 29.5$ (red), 
4465 galaxies from CANDELS/GOODS-S Deep with $H_{160} < 27$ (cyan), 6947 galaxies from CANDELS/GOODS-S Wide with $H_{160} < 26$ (blue), 
and 23090 galaxies from the UltraVISTA/COSMOS field with $i' < 26.0$ (yellow).}
\end{figure}

\subsection{Final Galaxy Sample}

The final galaxy sample consists of 36051 galaxies selected to lie in the redshift range $1.5 < z < 4.5$, 
and consists of: {\bf i)} 1549 galaxies from the HUDF with $H_{160} < 29.5$, {\bf ii)} 11412
galaxies from GOODS-S (comprising 4465 from CANDELS/GOODS-S Deep with $H_{160} < 27$ and 
6947 from CANDELS/GOODS-S Wide with $H_{160} < 26$), and {\bf iii)} 23090 galaxies from the UltraVISTA/COSMOS field with $i' < 26.0$.

The redshift distributions of these final HUDF, CANDELS/GOODS-S and UltraVISTA/COSMOS samples are shown in Fig.\,4.

Absolute magnitudes at UV rest-frame wavelengths $\lambda_{rest} = 1500$\,\AA\ and $\lambda_{rest} = 1700$\,\AA\ for use 
in the subsequent LF analyses were computed from the {\sc Le Phare} SED fits using a 100\,\AA\ top-hat 
synthetic filter centred at the appropriate rest wavelength.

\section{The Galaxy Luminosity Function}

Armed with redshifts and absolute UV magnitudes for over 35,000 galaxies in the redshift range $1.5 < z < 4.5$, we can 
now derive the rest-frame UV galaxy LF, exploring its form and evolution from $z = 2$ to $z = 4$.
To aid comparison with previous results in 
the literature, we derive LFs at both $\lambda_{rest} = 1500$\,\AA\ and $\lambda_{rest} = 1700$\,\AA\ as required,
but for our final calculations of the evolution of LF parameters, and the evolution of luminosity density (and hence star-formation rate density), 
we focus on 1500\,\AA\ at all redshifts.

In the first subsection below we outline the (straightforward) method we have adopted to determine the non-parametric binned form of the LF,
and then the parametric form (in this case the Schechter function).

We then present and discuss our results at various redshifts, in part to facilitate comparison with the literature. Specifically, we consider  
first the UV LF at 1500\,\AA\ in the redshift range $1.5<z<2.5$ ($z\simeq 2$) (in particular focussing on the faint-end slope, $\alpha$), 
before considering separately the 1500\,\AA\ LF at $1.5<z<2$ ($z\simeq 1.7$) and the 1700\,\AA\ LF at  $2<z<2.5$ ($z\simeq 2.2$). We then move on to determine  
the evolution of the 1500\,\AA\ galaxy LF over the redshift range $z\simeq 1.5-4.5$, in three redshift bins of width $\Delta z=1$ (i.e. 
corresponding to $z\simeq 2$, $z \simeq 3$ and $z \simeq 4$).

\begin{figure} 
\centering	
\includegraphics[width=0.47\textwidth]{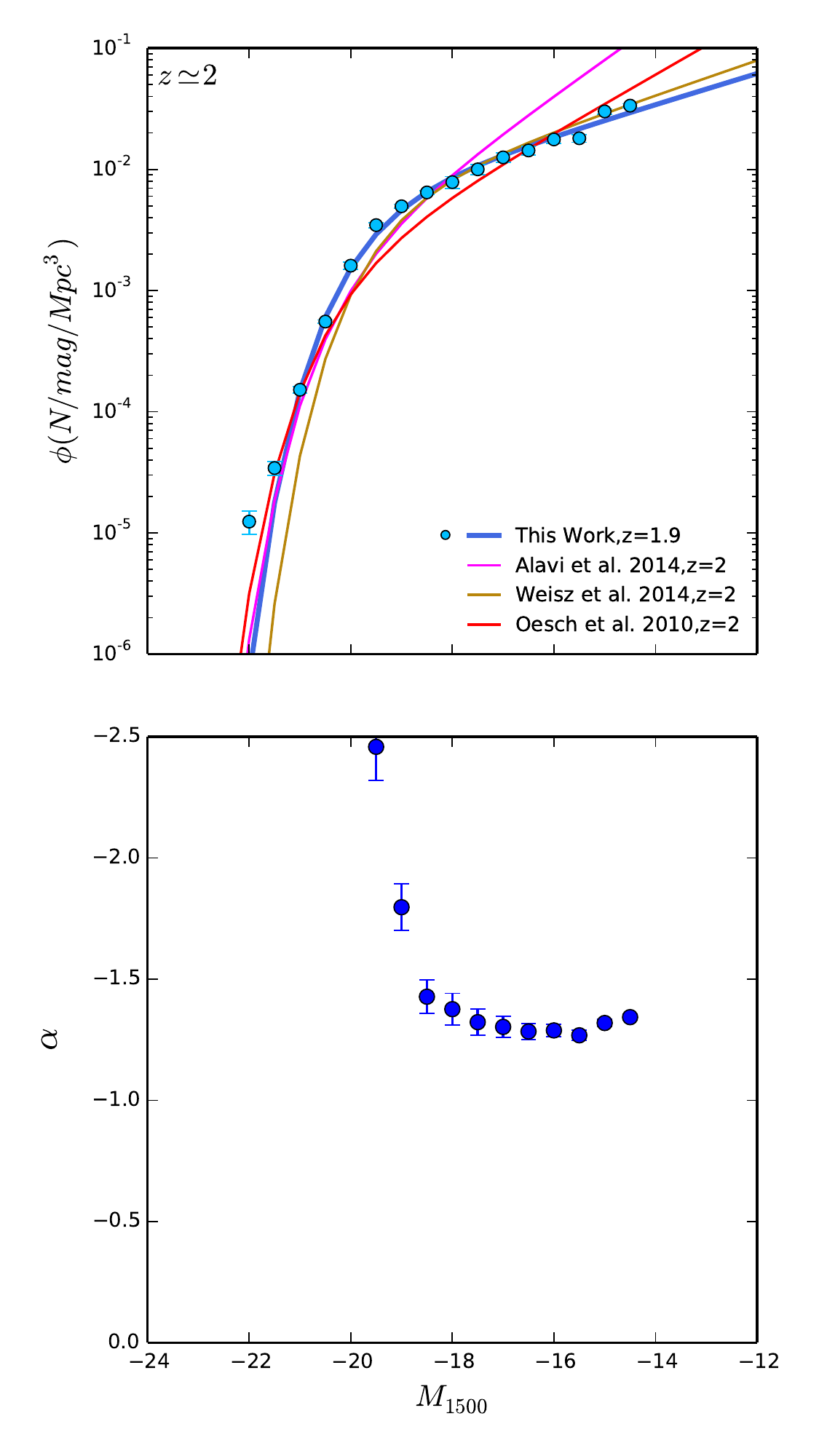}
\caption{The galaxy rest-frame UV LF at $z \simeq 2$. The upper panel shows 
the new 1500\,\AA\ LF 
as derived from our combined galaxy sample in the redshift 
range $1.5 < z < 2.5$. 
The blue circles with errors 
indicate the number densities from the $V_{max}$ estimator 
(see Table 2) and the blue line is our best-fitting Schechter function. The red and pink lines are the 
Schechter functions at $z \simeq 2$ reported  by Oesch et al. (2010) and Alavi et al. (2014) 
respectively, both of whom derived a much steeper faint-end slope at $z \simeq 2$. The 
orange line shows the $z \simeq 2$ LF as inferred by Weisz et al. (2014); this has a significantly  
shallower faint-end slope, in excellent agreement with the value 
of $\alpha$ deduced from our new determination (see Section 4.2.1). 
The lower panel shows how the fitted value of $\alpha$ 
depends on the limiting absolute magnitude down to which the 
fitting is performed.
It can be seen that the derived faint-end slope stabilises at $M_{1500} > -17$, settling to a secure 
and robust value of $\alpha = -1.32 \pm 0.03$.}
\end{figure}

\subsection{Method}

Various techniques can be ultilized to derive the LF, but here we have sufficiently extensive and dense coverage of the luminosity-redshift plane
to obtain a non-parametric estimate by applying the straightforward $V_{max}$ estimator (Schmidt 1968) given by:

\begin{equation} 
\centering
\phi(M)dM=\sum_{i}[\frac{1}{C(m_{i})V_{max,i}}]
\end{equation}

\noindent
where the sum is over all galaxies in the given redshift and absolute magnitude bin (chosen here 
to have a width of $\Delta{M}=0.5$\,mag), $V_{max}$ for each galaxy is set by the upper redshift 
limit of the bin unless the source drops out of the sample before that redshift is achieved, 
and $C$ is the completeness factor for each source. The errors on the derived number density 
in each bin are here assumed  to be Poissonian.

The completeness factor corrects for incompleteness
caused both by the fact that significant regions of the imaging are in practice inaccessible for high-redshift 
object selection (i.e. areas masked due to the presence of bright 
foreground galaxies or stars/diffraction spikes) and by photometric scatter (which obviously impacts most seriously 
on the faintest magnitude bins). This has been calculated by Monte Carlo source injection and retrieval 
simulations, and over most of the magnitude range in each sample transpires to be $\simeq 70$\% in the HUDF and 
CANDELS/GOODS-S fields, and $\simeq 60$\% in the UltraVISTA/COSMOS field. 

Obviously, within each of the three survey fields utilized here, incompleteness becomes more serious as the 
detection limit is approached, and the impact of photometric scatter becomes significant. However, in the present 
study the impact of this is minimal, as there is sufficient overlap between the regions of the luminosity-redshift 
plane covered by the different surveys that we can, for example, discard all seriously incomplete faint bins from the 
UltraVISTA/COSMOS survey in favour of the first well-sampled brighter bins from the CANDELS/GOODS-S survey (and similarly
ensuring the LF determination is dictated by the HUDF before CANDELS/GOODS-S becomes seriously incomplete).
At the very faint end we neglect all bins delivered by the HUDF in which photometric scatter results 
in a completeness $< 90$\%.

Finally, having derived the non-parametric LF from the combination of our three surveys, we fit 
the binned values with a Schechter function (Schechter 1976):  

\begin{equation} 
\centering
\phi(M)=0.4ln10\phi^{*}(10^{-0.4(M-M^{*})})^{\alpha+1}$e$^{-10^{-0.4(M-M^{*})}}
\end{equation}

\noindent
where $\phi^{*}$, $M^{*}$ and $\alpha$ indicate respectively the normalisation 
coefficient, the characteristic magnitude and the faint end slope of the LF, and derive confidence 
intervals on the parameters.

\begin{figure*}
\includegraphics[width=0.9\textwidth]{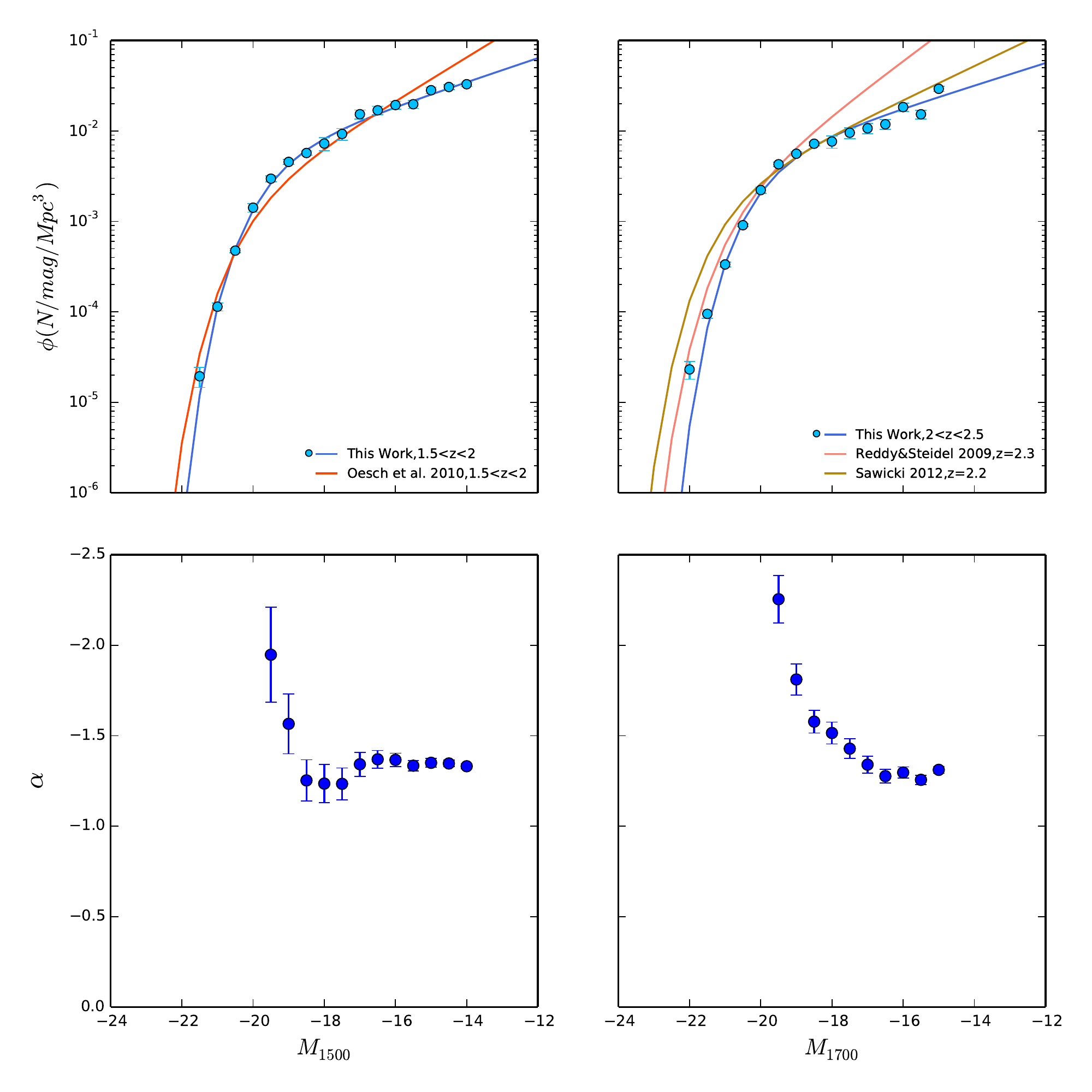}
\caption{Upper-Left: The 1500\,\AA\ LF as derived for galaxies in the redshift bin $1.5<z<2$. The 
blue data points show the number densities measured using the $V_{max}$ method and the solid blue line is our  
best-fitting Schechter function. The yellow solid line with the steeper faint-end slope is the 
best-fitting Schechter LF measured by Oesch et al. (2010) over the same photometric redshift range. 
Upper-Right: The 1700\,\AA\ LF derived over the photometric redshift 
range $2<z<2.5$. Again the blue data points show the binned number densities as 
derived from the $V_{max}$ method, while the solid blue line is our best-fitting Schechter 
function. The pink and yellow lines indicate, respectively, the Schechter function LFs 
derived by Reddy \& Steidel (2009) and Sawicki (2012), based on colour-selection. The 
lower panels again show how the fitted value of $\alpha$ 
depends on the limiting absolute magnitude down to which the 
fitting is performed.}
\end{figure*}

\subsection{The galaxy UV LF at ${\bf z\simeq 2}$}

\subsubsection{$1.5<z<2.5$}

We first derive a new measurement of the galaxy rest-frame UV LF at $z \simeq 2$, based on all galaxies 
in our combined sample with photometric redshifts in the range $1.5 < z < 2.5$. Fig.\,5 shows the resulting
LF at $\lambda_{rest} = 1500$\,\AA, including our best-fitting Schechter function. Here, bins brighter 
than $M_{1500} = -20$ are dominated by the UltraVISTA COSMOS sample, while at the faintest magnitudes 
the HUDF sample allows us to extend the UV LF down to $M_{1500} \simeq -14.5$, which is $\simeq 5$ mag fainter    
than achieved by Oesch et al. (2010). The extension of the $z \simeq 2$ LF to comparably 
faint magnitudes has only previously been reported by Alavi et al. 
(2014). However, this was only achieved with the aid of the gravitational lensing provided by the cluster 
Abell 1689, resulting in very small effective survey volumes and consequently much poorer S/N than achieved here.

The most striking result of our new $z \simeq 2$ LF determination, as shown in the upper panel of Fig.\,5, 
is that we find a much shallower faint-end slope ($\alpha = -1.32 \pm 0.03$) 
than reported by either Oesch et al. (2010) or Alavi et al. (2014) who found 
$\alpha = -1.60 \pm 0.51$ and $\alpha = -1.74 \pm 0.08$ respectively.
The lower panel in Fig.\,5 shows that an accurate measurement of $\alpha$ requires good sampling of the LF at magnitudes fainter 
than $M_{1500} \simeq -17$, after which the fitted value of $\alpha$ stabilises and yields a robust measurement.
It is therefore not surprising that Oesch et al. (2010) deduced an erroneously steep faint-end 
slope given the limited depth of the data utilised in that study. 

Our best-fitting values of the other Schechter parameters at $z \simeq 2$ are $M^*=-19.68 \pm 0.05$ and 
$\phi^*=7.02 \pm 0.66$ $(\times10^{-3} {\rm Mpc^{-3}mag^{-1}})$. Also shown in Fig.\,5 is the $z \simeq 2$ 
UV LF inferred by Weisz et al. (2014) from local-group galactic archaeology. Interestingly it is this 
`reverse engineered' LF which agrees best with our new direct determination, as Weisz et al. (2014) 
also infer a very similar, shallow faint-end slope of $\alpha = -1.36 \pm 0.11$.

\subsubsection{$1.5<z<2$ and  $2<z<2.5$}

To further facilitate comparison with previous studies, we next compute 
the UV LF within finer redshift bins, $1.5<z<2$ and $2<z<2.5$. We calculate the LF in the 
latter redshift bin at $\lambda_{rest} = 1700$\,\AA\ to simplify direct comparison 
with the results of Reddy \& Steidel (2009) and Sawicki (2012), both of whom calculated 
the LF at 1700\,\AA\ based on colour selection sampling an effective redshift window
$1.9<z<2.7$. 

Our results at $z \simeq 1.7$ and $z \simeq 2.2$ are shown in Fig.\,6. As in Fig.\,5 the upper panels 
show the binned LF and best-fitting Schechter function, while the lower panels show the derived 
value of faint-end slope, $\alpha$, as a function of the absolute magnitude limit down to 
which the fitting is performed. Again it can be seen that the derived value of $\alpha$ 
only stabilises at $M_{UV} > -17$, and that studies reaching only $M_{UV} \simeq -19$ are 
likely to yield an eroneously steep faint-end slope.

Our derived Schechter parameter values for the 1500\,\AA\ LF at $z \simeq 1.7$ 
are $M^* = -19.61 \pm 0.07$, $\phi^* = 6.81 \pm 0.81\,(\times10^{-3} {\rm Mpc^{-3}mag^{-1}})$, 
and $\alpha = -1.33 \pm 0.03$. In the left-hand panel of Fig.\,6 we also show the Schechter
function derived by Oesch et al. (2010) in the same redshift interval; it can be seen that
while the bright end is comparable, Oesch et al. (2010) inferred a much steeper faint-end slope of
$\alpha = -1.6 \pm 0.21$.

For the 1700\,\AA\ LF at $z \simeq 2.2$ we find
 $M^* = -19.99 \pm 0.08$, $\phi^* = 6.20 \pm 0.77\,(\times10^{-3} {\rm Mpc^{-3}mag^{-1}})$, 
and $\alpha = -1.31\pm0.04$. In the right-hand panel of Fig.\,6 we also show the Schechter
functions derived by Reddy \& Steidel (2009) and Sawicki (2012) from very similar redshift 
ranges. While our LF matches that derived by Reddy \& Steidel (2009) around the break, the faint-end 
slope derived by Reddy \& Steidel (2009) was clearly much steeper, with $\alpha =-1.73 \pm 0.07$.
The faint-end slope derived by Sawicki (2012) was somewhat shallower (although still steeper 
than our new derivation), but rather uncertain ($\alpha = -1.47 \pm 0.24$). Moreover, it can 
also be seen that the bright-end of the LF as derived by Sawicki (2012) also deviates 
significantly from our new results.

\begin{figure*}
\centering
\includegraphics[width=1\textwidth,height=0.25\textheight]{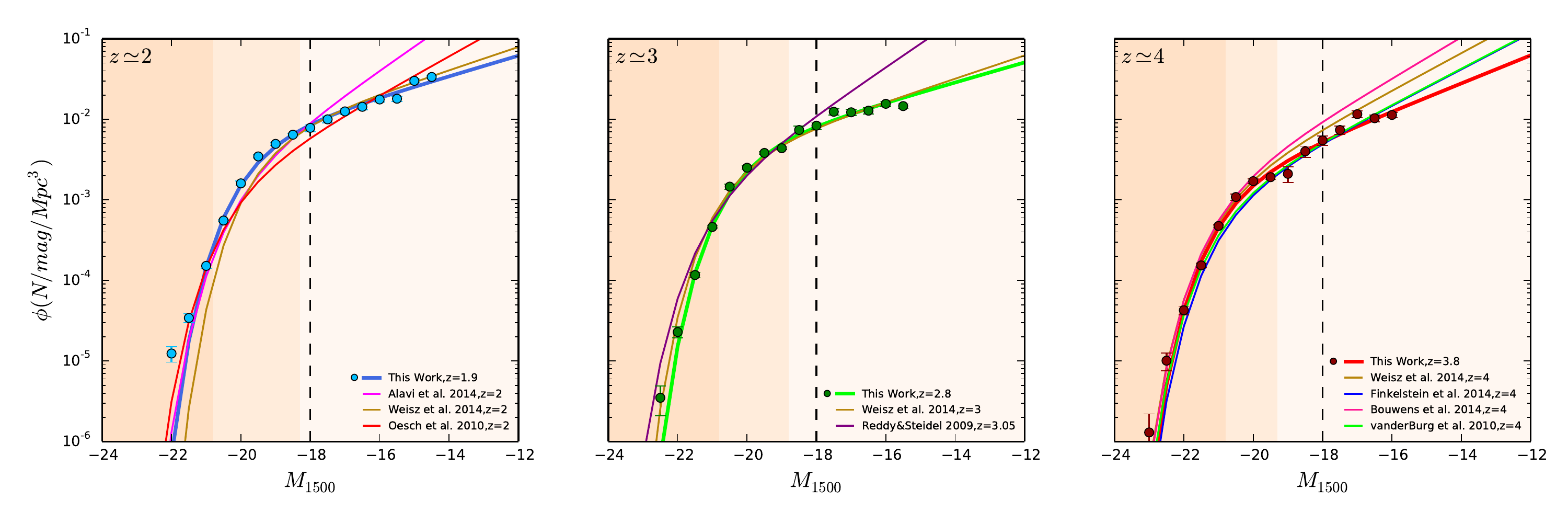}
\caption{Our new rest-frame UV (1500\,\AA) galaxy luminosity functions at 
$z \simeq 2$, 3 and 4. The (blue, green, red) data points indicate the 
values derived via the $V_{max}$ estimator, with the colour-matched solid
lines showing the best-fitting Schechter functions. The values corresponding to
the data points and their errors are tabulated in Table 2, while the values of the 
best-fitting Schechter parameters are given in Table 3, and plotted in Fig.\,8.
The vertical dashed line in each panel at $M_{1500} = -18$ is shown simply to 
indicate the typical absolute magnitude limit reached by previous studies 
at these redshifts, while the background shading indicates the absolute magnitude regimes in which the three different surveys (UltraVISTA/COSMOS, CANDELS/GOODS-S, HUDF) make the dominant contribution to our new measurement of the LF
at each redshift. 
For comparison purposes, at each redshift we also show several luminosity functions from the literature 
as indicated in the legend, and discussed in Section 4.3.}
\end{figure*}

\begin{table*}
	\centering
\caption{The rest-frame UV (1500\,\AA) galaxy luminosity functions 
at $z \simeq 2$, 3 and 4, as measured via the $V_{max}$ estimator; these values
are plotted in the three panels of Fig.\,7, and in Fig.\,9.}
\begin{tabular}{|l|l|l|l|}
\hline
  \multicolumn{1}{|c|}{$M_{1500}$} &
  \multicolumn{1}{c|}{$\phi$ ($z \simeq 2$) } &
  \multicolumn{1}{c|}{$\phi$ ($z \simeq 3$) } &
  \multicolumn{1}{c|}{$\phi$ ($z \simeq 4$) } \\
  & /${\rm Mpc^{-3}mag^{-1}}$ & /${\rm Mpc^{-3}mag^{-1}}$ & /${\rm Mpc^{-3}mag^{-1}}$\\
\hline
$-$23	&	$-$				&	$-$	                &	$0.000001 \pm 0.000000$	\\
$-$22.5	&	$-$				&	$0.000003 \pm 0.000001$	&	$0.000010 \pm 0.000002$	\\
$-$22	&	$0.000012 \pm 0.000027$		&	$0.000023 \pm 0.000004$	&	$0.000043 \pm 0.000005$	\\
$-$21.5	&	$0.000034 \pm 0.000045$		&	$0.000117 \pm 0.000008$	&	$0.000154 \pm 0.000010$	\\
$-$21	&	$0.000152 \pm 0.000094$		&	$0.000462 \pm 0.000016$	&	$0.000475 \pm 0.000017$	\\
$-$20.5	&	$0.000555 \pm 0.000181$		&	$0.001462 \pm 0.000107$	&	$0.001087 \pm 0.000096$	\\
$-$20	&	$0.001654 \pm 0.000124$		&	$0.002511 \pm 0.000140$	&	$0.001709 \pm 0.000120$	\\
$-$19.5	&	$0.003467 \pm 0.000165$		&	$0.003830 \pm 0.000173$	&	$0.001916 \pm 0.000127$	\\
$-$19	&	$0.004961 \pm 0.000197$		&	$0.004387 \pm 0.000185$	&	$0.002110 \pm 0.000467$	\\
$-$18.5	&	$0.006454 \pm 0.000225$		&	$0.007382 \pm 0.000838$	&	$0.004008 \pm 0.000644$	\\
$-$18	&	$0.007849 \pm 0.000869$		&	$0.008353 \pm 0.000892$	&	$0.005485 \pm 0.000753$	\\
$-$17.5	&	$0.010007 \pm 0.000981$		&	$0.012432 \pm 0.001088$	&	$0.007384 \pm 0.000874$	\\
$-$17	&	$0.012560 \pm 0.001099$		&	$0.012238 \pm 0.001079$	&	$0.016030 \pm 0.001095$	\\
$-$16.5	&	$0.001432 \pm 0.001173$		&	$0.012821 \pm 0.001105$	&	$0.010337 \pm 0.001034$	\\
$-$16	&	$0.017660 \pm 0.001303$		&	$0.015599 \pm 0.001216$	&	$0.013510 \pm 0.000850$	\\
$-$15.5	&	$0.018052 \pm 0.001317$		&	$0.014625 \pm 0.000971$	&	$-$			\\
$-$15	&	$0.030077 \pm 0.001505$		&	$-$			&	$-$			\\
$-$14.5	&	$0.033572 \pm 0.001251$		&	$-$			&	$-$			\\		

\hline\end{tabular}

\end{table*}

\subsection{The galaxy UV LF at ${\bf z \simeq 3}$ and ${\bf z \simeq 4}$}

We now extend our study of the galaxy UV LF out to higher redshift, considering also 
the redshift bins $2.5 < z < 3.5$ and $3.5 < z < 4.5$, in order to explore how the LF 
evolves over the crucial redshift range $z \simeq 2 - 4$.

We have focussed on 1500\,\AA\, and our results are shown in Fig.\,7, alongside the $z \simeq 2$ 1500\,\AA\ LF 
which was shown in Fig.\,5. 
The binned number densities derived from the $V_{max}$ method at $z \simeq 2$, 3 and 4, as shown in Fig.\,7, are tabulated 
in Table~1, and our derived best-fitting Schechter function parameter values at all three redshifts are included 
in Table~2 (along with various values from the literature, as discussed further in Section 5).

Again, for comparison, in Fig.\,7 we also overplot other recent determinations of the 1500\,\AA\ LF at these redshifts.
Our derived LFs at $z \simeq 3$ and $z \simeq 4$ appear to agree reasonably well with previous measurements
around the break luminosity but, as at $z \simeq 2$, we find a shallower faint-end slope, $\alpha$, than most 
previous studies; again, we agree best with the results inferred from the local Universe by Weisz et al. (2014)
(although we note that, at $z \simeq 3$, Weisz et al. (2014) have adopted the bright end of the 1500\,\AA\ LF 
given by Reddy \& Steidel (2009)). At $z \simeq 4$ we find a shallower faint-end slope than most previous 
studies (i.e. $\alpha = -1.43 \pm 0.04$), but $\phi^*$ and $M^*$ are in excellent agreement with the
results of Bouwens et al. (2014) (see Table~3).

In Fig.\,8 we plot our derived Schechter parameter values, with 1-$\sigma$ and 2-$\sigma$ 
single-parameter confidence intervals, for $z \simeq 2$, $z \simeq 3$ and $z \simeq 4$, while in Fig.\,9 we 
overplot the 1500\,\AA\ LFs at these three redshifts. These plots highlight evolutionary trends in the 
UV LF over this key redshift range, which we discuss further below in Section 5.

\section{Discussion and Implications}

\begin{figure} 
\centering
\includegraphics[width=0.47\textwidth]{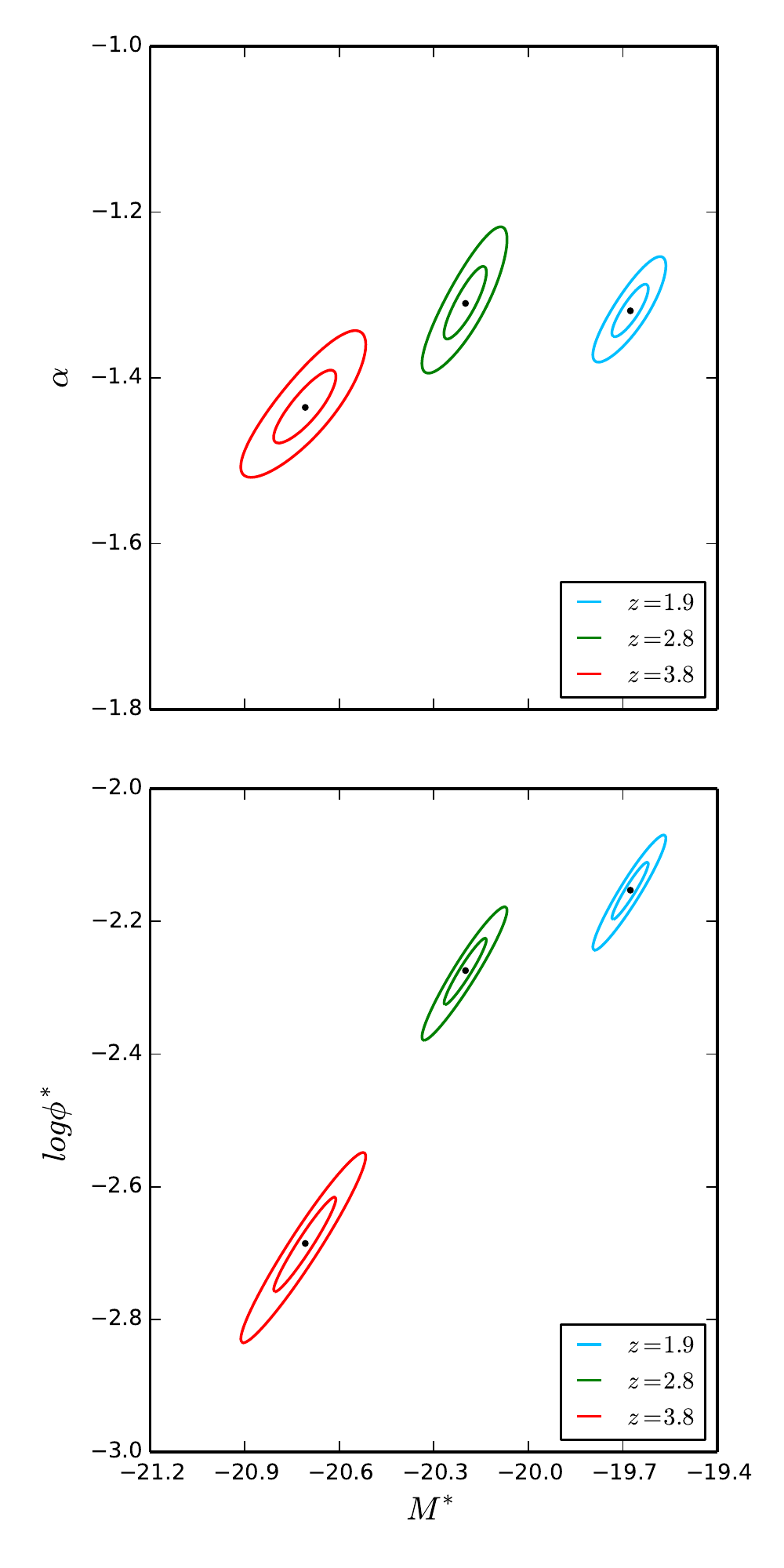}
\caption{Our derived best-fitting Schechter function parameter values at $z \simeq 2$, 3, and 4, with
the associated 1-$\sigma$ and 2-$\sigma$ single-parameter confidence regions 
(corresponding to $\Delta \chi^2 = 1, 4$ after minimizing over the other parameter). As can be seen from 
the upper panel, $\alpha$ remains shallow, steepening little if at all over the redshift range. 
The lower panel shows that $\phi^*$ drops gradually from $z \simeq 2$ to $z \simeq 3$, but then
falls by a factor $\simeq 2.5$ by $z \simeq 4$. However, this drop in number density is offset 
by a steady brightening in $M^*$ by $\simeq 1$\,mag. from $z \simeq 2$ to $z \simeq 4$, to the extent that
UV luminosity density peaks at $z \simeq 2.5 - 3$; see Fig.\,11.}
\end{figure}

\begin{figure} 
\centering
\includegraphics[width=0.47\textwidth]{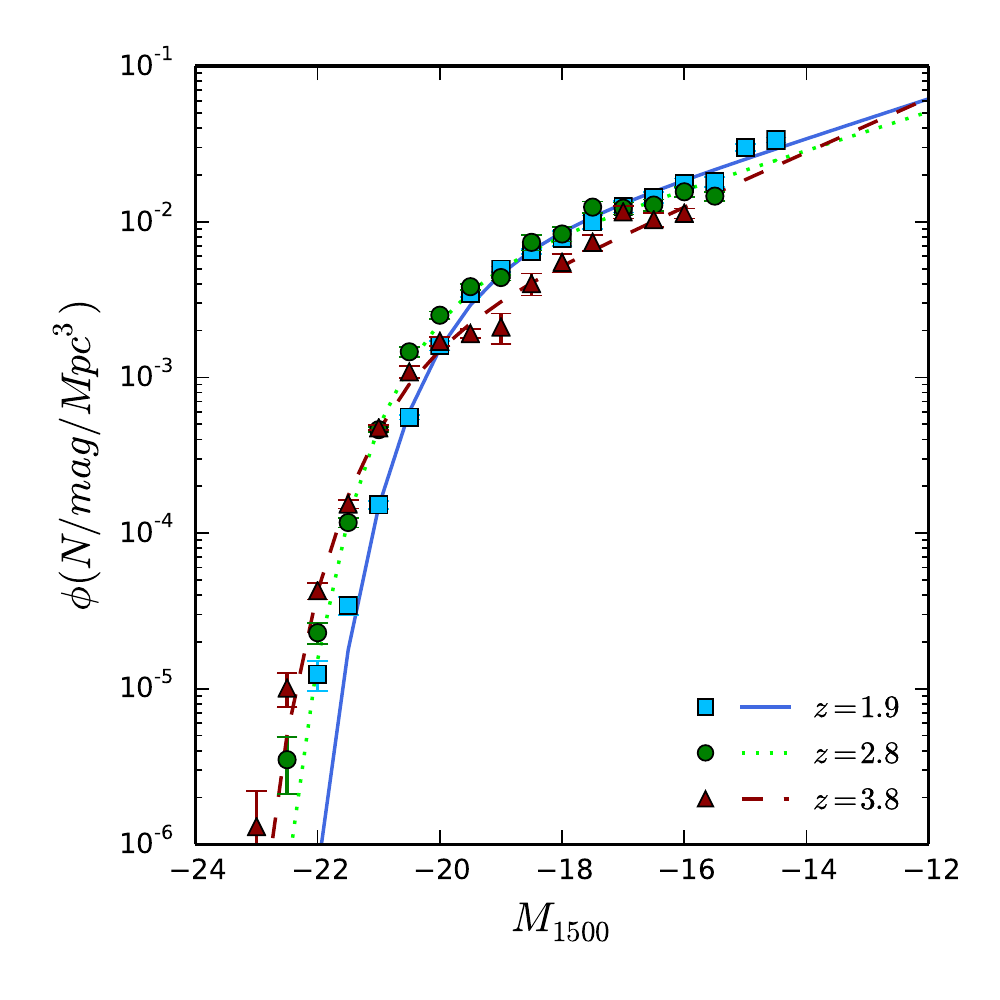}
\caption{The rest-frame UV (1500\AA) galaxy luminosity functions presented in Fig.\,7, now overlaid 
to show the form of the evolution from $z \simeq 2$ to $z \simeq 4$. As is also clear 
from the Schechter function parameter values plotted in Fig.\,8, it can be seen that 
i) the faint-end slope is little changed over this redshift range, ii) normalization drops 
only slightly between $z \simeq 2$ and $z \simeq 3$
but then drops by a factor $\simeq 2.5$ by $z \simeq 4$, and iii) the LF brightens steadily
by $\simeq 1$\,mag. from $z \simeq 2$ to $z \simeq 4$.}
\end{figure}

\subsection{The evolution of the LF from ${\bf z \simeq 2}$ to ${\bf z \simeq 4}$}

As can be seen in Fig.\,8 and Fig.\,9, while the LF displays relatively little evolution between
$z \simeq 2$ and $z \simeq 3$, there is a clear drop in $\phi^*$ (by a factor $\simeq 2.5$)
between $z \simeq 3$ and $z \simeq 4$. We also find, however, that $M^*$ brightens steadily 
over this redshift range, by $\simeq 1$\,mag.. In terms of luminosity density, this brightening 
more than offsets the decline in $\phi^*$ up until $z \simeq 3$, but by $z \simeq 4$ the more dramatic 
drop in $\phi^*$ dominates the evolution, and luminosity density undoubtedly declines significantly 
by $z \simeq 4$.

The modest evolution in $\phi^*$ and the $\simeq 0.5$ mag. 
brightening in $M^*$ seen between $z \simeq 2$ and $z \simeq 3$ is very similar to the
evolution reported by Reddy \& Steidel (2009), with the main difference being that 
our $\phi^*$ values are systematically higher, and our $M^*$ values systematically dimmer 
due (at least in part) to our significantly shallower best-fitting faint-end slope $\alpha$.
This is made clear in Fig.\,10, where we place our results in the context of 
several other recent studies. Here it can be seen that our inferred values of $\phi^*$ 
at $z \simeq 2$ and $z \simeq 3$ are noticeably higher than derived in nearly all 
previous studies, while our inferred values of $M^*$ are 
therefore (unsurprisingly) somewhat lower. Our results in fact agree best 
with those recently derived by Weisz et al. (2014), who attempted to 
reconstruct the form of the UV LF out to $ z \simeq 5$ from the properties 
(including star-formation histories) of the galaxies in the local 
group. We speculate that this agreement perhaps reflects the fact 
that our own study and that undertaken by Weisz et al. (2014) are 
the only studies to date which have probed to the depths required to properly determine 
the faint-end slope (in fact Weisz et al. (2014) reach down to $M_{UV} \simeq -5$ at $z \simeq 0$), 
with an inevitable resulting impact on the inferred best-fitting values of the 
other two Schechter parameters. While this agreement is interesting, and arguably impressive, 
we note that, unsurprisingly, the uncertainties in our
parameter values are much smaller than those presented by Weisz et al. (2014).

Interestingly, the evolution of $\phi^*$ and $M^*$ derived here over the redshift range
$z \simeq 2 - 4$, is also very similar to that recently derived for the 
evolving emission-line galaxy LFs by Khostovan et al. (2015) (although their results were 
derived by locking the value of the faint-end slope, due to the lack of data of sufficient depth
to constrain it).

\subsection{Evolution up to ${\bf z\simeq 8}$}

In Fig.\,10 we also attempt to place our findings in the wider context of the 
results derived from a number of recent studies 
of the UV LF extending out to $z \simeq 8$. It can be seen that 
the level of agreement is in fact better at $z \simeq 4$ than at $ z \simeq 2$ or $z \simeq 3$.
The solid black lines in each panel of Fig.\,10 show simple parametric 
fits to the data (i.e. to the published Schechter parameter values), to
illustrate the overall evolutionary trend in each parameter as would be 
derived from the literature. These curves are meant to guide the eye, and are not meant 
to indicate our best estimate of true parameter evolution; indeed our new, more accurate 
determinations of $\phi^*$ and $M^*$ at $z \simeq 2 - 3$ clearly differ significantly 
from the literature average (for the well-understood reasons described above).
Nevertheless, the evolutionary trend in $\alpha$ indicated by the simple 
straight-line fit 
shown in the bottom panel serves to re-emphasise how well our derived shallow faint-end slopes 
agree with the Weisz et al. (2014) results, and also
shows that such values are in fact in very reasonable agreement with the 
general trend of a gradual steepening from $\alpha \simeq -1.2$ at $z \simeq 0$ to 
$\alpha \simeq -2$ by $z \simeq 8$.

\begin{figure*}
\includegraphics[width=1\textwidth,height=0.75\textheight]{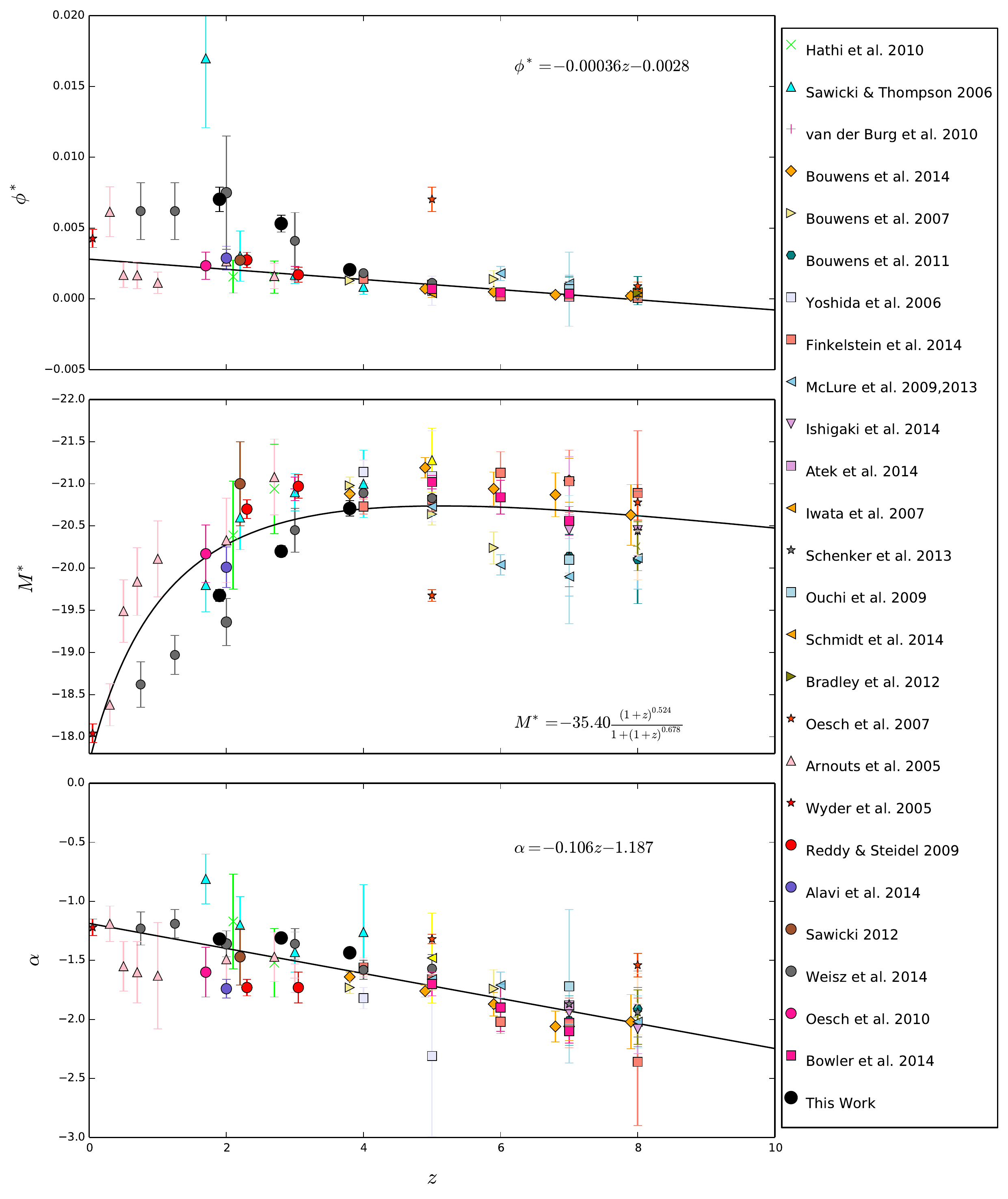}
\caption{A compilation of the derived Schechter function parameter values 
for the UV galaxy LF over the redshift range $z \simeq 0$ to $z \simeq 8$, 
placing the new results derived in this paper, and the other results tabulated in Table 3, 
into the wider context of virtually all of cosmic time. Our own results, with error bars
(see Fig.\,8), are shown by the large black points, with other results from the literature plotted as indicated in the legend. 
The solid black line in each panel is a simple parametric 
fit to the data, plotted to illustrate the overall evolutionary trend in each parameter as inferred from 
the literature. The evolution of $\phi^*$ and $M^*$ from $z \simeq 2$ to $z \simeq 4$ 
deduced in the present study is somewhat more dramatic than seen in most pre-existing 
direct studies of the LF at these redshifts, in fact agreeing best with the values 
inferred by Weisz et al. (2014) from galactic archaeology of the 
local group. The bottom panel again re-emphasises 
how well our derived shallow faint-end slopes agree with the Weisz et al. (2014) results, but also
shows that such values are in fact in reasonable agreement with the general trend of a gradual 
steepening from $\alpha \simeq -1.2$ at $z \simeq 0$ to 
$\alpha \simeq -2$ by $z \simeq 8$.}
\end{figure*}

\begin{table*}
\centering
\caption{A compilation of the derived Schechter function parameter values 
for the UV galaxy LF over the redshift range $z \simeq 1.7$ to $z \simeq 4.0$, 
placing the new results derived in this paper in the context of results presented 
in the literature over the last 10 years. The values tabulated here are included in Fig.\,10}
\begin{tabular}[width=0.5\textwidth]{|l|l|l|l|l|l|l|}
\hline
  \multicolumn{1}{|c|}{Source} &
  \multicolumn{1}{c|}{$z$} &
  \multicolumn{1}{c|}{$\lambda_{rest}$\,/\,\AA} &
  \multicolumn{1}{c|}{$M^{*}$} &
  \multicolumn{1}{c|}{$\phi^{*}$} &
  \multicolumn{1}{c|}{$\alpha$} \\
\\
\hline
  \textbf{This Work} & 1.7 & 1500          & $-19.61\pm0.07$ & $0.00681\pm0.00081$ & $-1.33\pm0.03$ \\
  Oesch et al. (2010) & 1.7 & 1500         & $-20.17\pm0.34$ & $0.00234\pm0.00096$ & $-1.60\pm0.21$ \\
  Sawicki \& Thomson (2006) & 1.7 & 1700   & $-19.80\pm0.32$ & $0.01698\pm0.00489$ & $-0.81\pm0.21$ \\
  Hathi et al. (2010) & 1.7 & 1500         & $-19.43\pm0.36$ & $0.00217\pm0.00077$ & $-1.27\pm0.00$ \\
  \textbf{This Work} & 1.9 & 1500          & $-19.68\pm0.05$ & $0.00702\pm0.00066$ & $-1.32\pm0.03$ \\
  Oesch et al. (2010)& 1.9 & 1500          & $-20.16\pm0.52$ & $0.00219\pm0.00123$ & $-1.60\pm0.51$ \\
  Arnouts et al. (2005) & 2.0 & 1500       & $-20.33\pm0.50$ & $0.00265\pm0.00020$ & $-1.49\pm0.24$ \\
  Alavi et al. (2014) & 2.0 & 1500         & $-20.01\pm0.24$ & $0.00288\pm0.00084$ & $-1.74\pm0.08$ \\
  Weisz et al. (2014) & 2.0 & 1500         & $-19.36\pm0.28$ & $0.00750\pm0.00400$ & $-1.36\pm0.11$ \\
  Hathi et al. (2010) & 2.1 & 1500         & $-20.39\pm0.64$ & $0.00157\pm0.00115$ & $-1.17\pm0.40$ \\
  Sawicki \& Thomson (2006) & 2.2 & 1700   & $-20.60\pm0.38$ & $0.00301\pm0.00176$ & $-1.20\pm0.24$ \\
  Sawicki (2012) & 2.2 & 1700              & $-21.00\pm0.50$ & $0.00274\pm0.00024$ & $-1.47\pm0.24$ \\
  \textbf{This Work} & 2.25 & 1500         & $-19.71\pm0.07$ & $0.00759\pm0.00088$ & $-1.26\pm0.04$ \\
  \textbf{This Work} & 2.25 & 1700         & $-19.99\pm0.08$ & $0.00620\pm0.00077$ & $-1.31\pm0.04$ \\
  Reddy \& Steidel (2009) & 2.3 & 1700     & $-20.70\pm0.11$ & $0.00275\pm0.00054$ & $-1.73\pm0.07$ \\
  Arnouts et al. (2005) & 2.7 & 1500       & $-21.08\pm0.45$ & $0.00162\pm0.00090$ & $-1.47\pm0.21$ \\
  Hathi et al. (2010) & 2.7 & 1500         & $-20.94\pm0.53$ & $0.00154\pm0.00114$ & $-1.52\pm0.29$ \\
  \textbf{This Work} & 2.8 & 1500          & $-20.20\pm0.07$ & $0.00532\pm0.00060$ & $-1.31\pm0.04$ \\
  Arnouts et al. (2005) & 3.0 & 1500       & $-21.07\pm0.15$ & $0.00140\pm0.00000$ & $-1.60\pm0.13$ \\
  Weisz et al. (2014) & 3.0 & 1500         & $-20.45\pm0.26$ & $0.00410\pm0.00200$ & $-1.36\pm0.13$ \\
  Sawicki \& Thomson (2006) & 3.0 & 1700   & $-20.90\pm0.22$ & $0.00167\pm0.00013$ & $-1.43\pm0.17$ \\
  van der Burg et al. (2010) & 3.0 & 1600  & $-20.94\pm0.14$ & $0.00179\pm0.00051$ & $-1.65\pm0.12$ \\
  Reddy \& Steidel (2009) & 3.05 & 1700    & $-20.97\pm0.14$ & $0.00171\pm0.00053$ & $-1.73\pm0.13$ \\
  \textbf{This Work} & 3.8 & 1500          & $-20.71\pm0.10$ & $0.00206\pm0.00033$ & $-1.43\pm0.04$ \\
  Bouwens et al. (2014) & 3.8 & 1600       & $-20.88\pm0.08$ & $0.00197\pm0.00034$ & $-1.64\pm0.04$ \\
  Bouwens et al. (2007) & 3.8 & 1600       & $-20.98\pm0.10$ & $0.00130\pm0.00020$ & $-1.73\pm0.05$ \\
  Weisz et al. (2014) & 4.0 & 1500         & $-20.89\pm0.11$ & $0.00182\pm0.00010$ & $-1.58\pm0.08$ \\
  Sawicki \& Thomson (2006) & 4.0 & 1700   & $-21.00\pm0.40$ & $0.00085\pm0.00021$ & $-1.26\pm0.40$ \\
  van der Burg et al. (2010) & 4.0 & 1600  & $-20.84\pm0.09$ & $0.00136\pm0.00023$ & $-1.56\pm0.08$ \\
  Finkelstein et al. (2015) & 4.0 & 1500   & $-20.73\pm0.09$ & $0.00141\pm0.00021$ & $-1.56\pm0.06$ \\
  Yoshida et al. (2006) & 4.0 & 1500        & $-21.14\pm0.14$ & $0.00146\pm0.00041$ & $-1.82\pm0.09$ \\
\hline\end{tabular}

\end{table*}

\subsection{Luminosity Density}

We finish by considering the evolution of UV luminosity density, $\rho_{UV}$,  
inferred from our LF determinations over the key redshift regime $z \simeq 2 - 4$.
While a full determination of star-formation rate density evolution also requires 
accounting for the substantial impact of dust obscuration, the luminosity-weighted
integral of the UV LF does provide an important measurement of the 
unobscured star-formation rate density at each redshift.

The results of this calculation are presented in Table 4 and plotted in Fig.\,11.
Here we have deliberately performed all calculations using the LFs determined 
at $\lambda_{rest} = 1500$\,\AA\ to enable unbiased comparison of the derived values 
at each redshift. Table 4 includes results calculated at $z \simeq 2$, 3 and 4, and 
also provides results for narrower redshift bins at $z \simeq 1.7$ ($1.5 < z < 2.0$)
and $z \simeq 2.25$ ($2.0 < z < 2.5$) to enable checking of the trend within the 
$z \simeq 2$ redshift bin. We also provide results integrated down to different limiting 
absolute magnitudes: $M_{1500} = -17.7$, 
$M_{1500} = -15$, and $M_{1500} = -10$. The results to $M_{1500} = -17.7$ are 
given for ease of comparison with many existing studies, while the convergence
seen at the deeper limits shows that,
because our derived faint-end slopes are fairly 
flat, relatively little additional luminosity density is contributed 
by the faintest galaxies; Fig.\,11 shows that $\rho_{UV}$ has essentially converged by 
$M_{1500} \simeq -15$. 
Regardless of the chosen integration limit, our results indicate that UV luminosity
density (and hence unobscured star-formation rate density) peaks at 
$z \simeq 2.5 - 3$, when the Universe was $\simeq 2.5$\,Gyr old.

The formal uncertainties indicated by the error bars in Fig.\,11 are fairly small, both because 
the integral of the LF is better constrained than the (somewhat degenerate) Schechter parameters, and because, 
with such a flat faint-end slope, uncertainties in $\alpha$ only have a minor effect on the 
luminosity-weighted integral. 
In practice, therefore, the true uncertainties are likely to be dominated 
by cosmic variance (although luminosity density 
is clearly less affected by cosmic variance uncertainties than, for example, bright galaxy number counts).

Finally, in Fig.\,12 we show our derived UV luminosity densities (integrated down to $M_{1500} = -17.7$) 
in the context of other recent determinations at comparable redshifts, and recent measurements extending to $ z \simeq 9$. 
Our new results are more accurate than previous determinations, but in 
generally good agreement with existing results at $z \simeq 3$ and $z \simeq 4$. At $z \simeq 2$
our new result lies at the low end of the (widely discrepant) previously reported measurements, but is in fact 
still higher than the recent estimate provided by Alavi et al. (2014). Even allowing for cosmic variance,
the basic conclusion that UV luminosity density peaks at $z \simeq 2.5 - 3$ appears secure.

\begin{table*}
\centering
\caption{The rest-frame UV (1500\AA) luminosity densities as derived from 
our UV LFs from $z \simeq 1.7$ to $z \simeq 4$, with the luminosity-weighted 
integral performed down to 
three different magnitude limits. Because our derived 
faint-end slopes are fairly 
flat, relatively little additional luminosity density is contributed 
by the faintest galaxies,
and Fig.\,11 shows that $\rho_{UV}$ has essentially converged by 
$M_{1500} \simeq -15$. 
Regardless of the chosen integration limit, it seems clear that UV luminosity
density (and hence unobscured star-formation density) peaks at $z \simeq 2.5 - 3$, when the Universe was $\simeq 2.5$\,Gyr old. The values given here
are plotted in Fig.\,11.}
\begin{tabular}{|c|c|c|c|}
\hline

  \multicolumn{1}{c|}{$z$} &
  \multicolumn{1}{c|}{$\rho_{uv} / {\rm 10^{26}\,ergs\,s^{-1} Hz^{-1} Mpc^{-3}}$} &
  \multicolumn{1}{c|}{$\rho_{uv} / {\rm 10^{26}\,ergs\,s^{-1} Hz^{-1} Mpc^{-3}}$} &
  \multicolumn{1}{c|}{$\rho_{uv} / {\rm 10^{26}\,ergs\,s^{-1} Hz^{-1} Mpc^{-3}}$} \\
  & $M_{limit}= - 10$ &  $M_{limit}= -15$ &  $M_{limit}= - 17.7$ \\
 \\
\hline
1.7  & $2.79 \pm 0.04$ & $2.62 \pm 0.05$ & $1.91 \pm 0.04$\\
1.9  & $3.01 \pm 0.04$ & $2.84 \pm 0.06$ & $2.11 \pm 0.07$\\
2.25 & $3.13 \pm 0.04$ & $3.00 \pm 0.06$ & $2.33 \pm 0.07$\\
2.8  & $3.64 \pm 0.01$ & $3.50 \pm 0.04$ & $2.86 \pm 0.07$\\
3.8  & $2.70 \pm 0.03$ & $2.56 \pm 0.05$ & $2.09 \pm 0.07$\\
\hline
\end{tabular}
\end{table*}

\begin{figure} 
\centering
\includegraphics[width=0.47\textwidth]{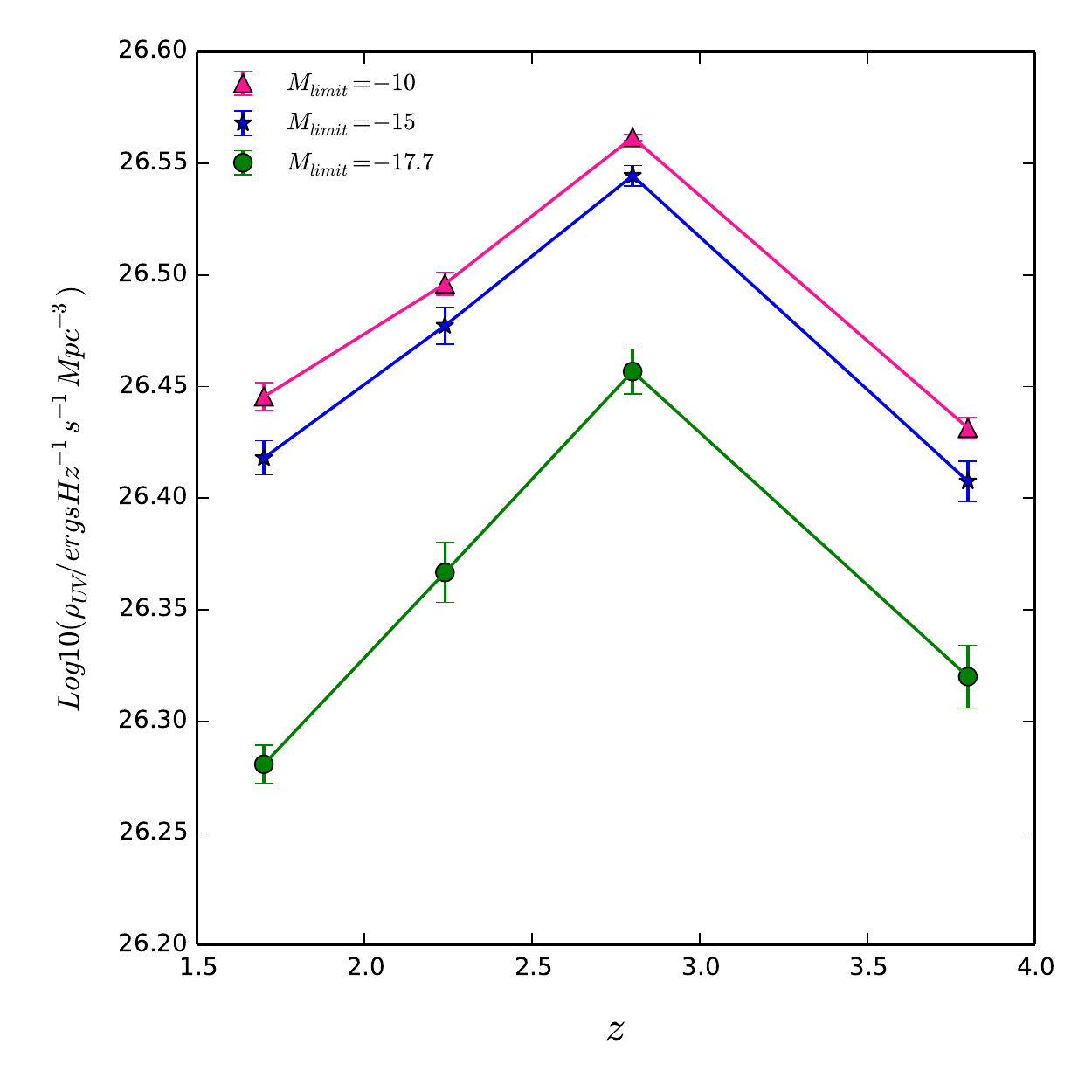}
\caption{The rest-frame UV (1500\AA) luminosity densities as derived from 
our UV LFs from $z \simeq 1.7$ to $z \simeq 4$, with the luminosity-weighted 
integral performed down to 
three different absolute magnitude limits: $M_{1500} = -17.7$ 
(green circles), $M_{1500} = -15$ (blue stars), and 
$M_{1500} = -10$ (pink triangles). The results to $M_{1500} = -17.7$ are 
shown for ease of comparison with many existing studies, while the convergence
seen at the deeper limits shows that,
because our derived faint-end slopes are relatively 
flat, relatively little additional luminosity density is contributed 
by the faintest galaxies (i.e. 
the plot shows that $\rho_{UV}$ has essentially converged by 
$M_{1500} \simeq -15$). 
Regardless of the chosen integration limit, it seems clear that UV luminosity
density (and hence unobscured star-formation density) peaks at $z \simeq 2.5 - 3$, when the Universe was $\simeq 2.5$\,Gyr old. 
The values plotted here
are tabulated in Table 4.}
\end{figure}

\begin{figure} 
\centering
\includegraphics[width=0.47\textwidth]{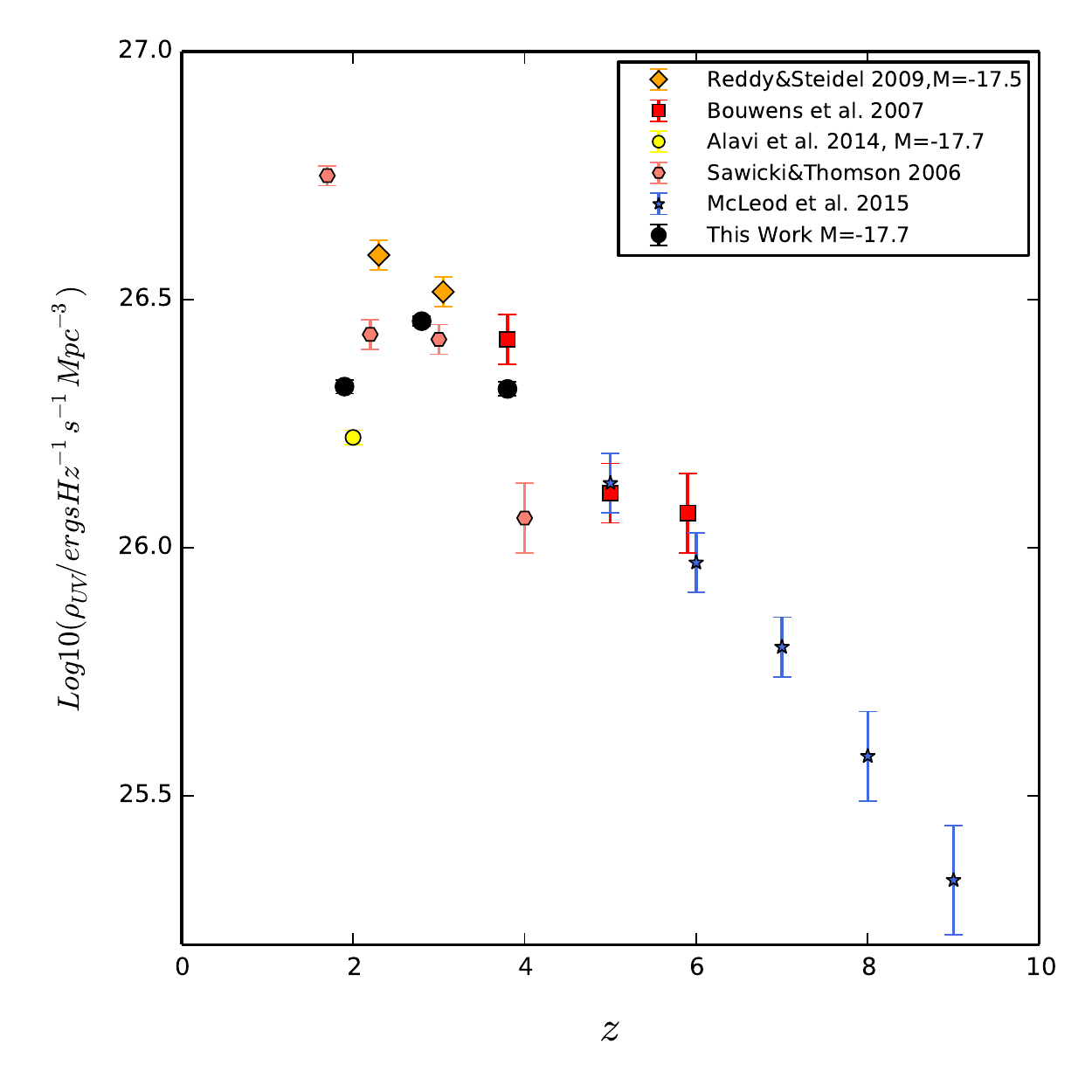}
\caption{Our derived UV (1500\AA) luminosity density values at $z \simeq 2$, 3 and 4, 
with integration performed down to $M_{1500} = -17.7$, are here plotted as the black points,
and compared to results of similar calculations performed by other authors 
as indicated in the legend. Our results are more accurate than previous determinations, but in 
generally good agreement with existing results at $z \simeq 3$ and $z \simeq 4$. At $z \simeq 2$
our new result lies at the low end of the (widely discrepant) previously reported measurements.}
\end{figure}

Of course, the precise epoch at which cosmic star-formation rate density 
reached a peak depends on the evolution of the correction for dust obscuration. At 
$z \simeq 2$ a number of arguments indicate that this correction involves 
scaling the raw UV luminosity density by a factor of $\simeq 4 - 5$ (e.g. 
Reddy \& Steidel 2009; Burgarella et al. 2013; Madau \& Dickinson 2014), 
but whether this correction evolves significantly between $z \simeq 3$ and $z \simeq 2$ 
remains as yet unclear. 

Recent reviews of cosmic star-formation history based on data compilations 
have generally favoured a peak in cosmic star-formation rate density at $z \simeq 2$ (e.g. Behroozi, Weschler \& Conroy 2013; 
Madau \& Dickinson 2014) but at least some recent studies (e.g. radio: Karim et al. 2011; far-infrared: 
Burgarella et al. 2013; emission-line: Khostovan et al. 2015) favour a peak nearer $z \simeq 3$. 
The latter scenario is more obviously consistent with the new UV results presented here, but a definitive 
answer awaits more direct measurements of dust-enshrouded star-formation at $z \simeq 2 - 4$ 
from forthcoming deep sub-mm/mm surveys with SCUBA-2 on the JCMT and the Atacama Large Millimeter Array.

\section{Conclusion}

We have exploited the high dynamic range provided by combining the 
Hubble Ultra Deep Field (HUDF), CANDELS/GOODS-South, and UltraVISTA/COSMOS surveys to derive 
a new, robust measurement of the evolving rest-frame ultraviolet
galaxy luminosity function over the key redshift range from $z \simeq 2$ to $z \simeq 4$. 

The unparalleled multi-frequency photometry available in this survey 
`wedding cake', combined with the (relative) wealth of deep optical and near-infrared spectroscopy
in these fields, has enabled us to derive accurate photometric redshifts for 
$\simeq 95$\% of the galaxies in the combined survey (with a reliability and accuracy 
that are competitive with the very best achieved to date, as verified in Appendix A).

This has then enabled us to assemble robust and complete galaxy samples within redshift 
slices at $z \simeq 2$, 3 and 4, facilitating a new determination of the 
form and evolution of the UV galaxy LF, that probes $\simeq 3 - 4$ 
magnitudes fainter than previous (unlensed) surveys at $z \simeq 2 - 3$, and does not rely on potentially 
incomplete colour-colour selection techniques. The SED fitting undertaken to determine the photometric redshifts
has also allowed us to determine accurate rest-frame UV absolute magnitudes ($M_{1500}$ or $M_{1700}$, as required
for comparison with previous results).

Our new determinations of the UV LF extend from $M_{1500} \simeq -22$ (AB mag) 
down to $M_{1500}$\,=\,$-$14.5, $-$15.5 and $-$16 at $z\simeq$\,2, 3 and 4 respectively.
Fitting a Schechter function to the LF data as determined from the $V_{max}$ estimator, 
at $z \simeq 2 - 3$ reveals a much shallower faint-end slope ($\alpha = -1.32 \pm 0.03$) 
than the steeper values ($\alpha \simeq -1.7$) reported by Reddy \& Steidel (2009) or by 
Alavi et al. (2014) (who utilised gravitional lensing to help sample the faint end of the LF).
By performing the Schechter function fitting down to differing limiting magnitudes, we show that
our measurement of faint-end slope is robust (i.e. the inferred value plateaus/converges 
at $M_{1500} > -17$). By $z \simeq 4$ the faint-end slope has steepened slightly, to $\alpha = -1.43 \pm 0.04$.
Although these values are significantly shallower than the aforementioned pre-existing estimates,
we find they are in excellent agreement with the values recently inferred by Weisz et al. (2014) 
(from galactic archaeology of the local group), and are in fact
consistent with the overall evolutionary trend in $\alpha$ from $z = 0$ to $z = 8$, as gleaned 
from a review of the literature. 

Analysis of the other best-fitting Schechter function parameters reveals that our derived number density 
normalization, $\phi^*$, is higher than nearly all previous estimates at $z \simeq 2$ (except, again, Weisz et al. 
2014), declines only slightly by $z \simeq 3$, and then drops by a factor $\simeq 2.5$ to $z \simeq 4$
(where our value agrees well with most previous measurements). Meanwhile, this drop in number density is offset 
by a steady brightening in $M^*$ by $\simeq 1$\,mag. from $z \simeq 2$ to $z \simeq 4$, to the extent that
UV luminosity density does not drop significantly until the negative density evolution
takes over and dominates beyond $z \simeq 3$.

Finally, we have compared our new UV LF determinations, and the resulting inferred evolution of 
UV luminosity density ($\rho_{UV}$), with results from a range of previous studies 
extending from $z \simeq 0$ out to $z \simeq 9$. Because our new measurements 
yield fairly flat faint-end slopes, our estimates of $\rho_{UV}$ are relatively robust; they  
have essentially converged by $M_{UV} \simeq -15$, and are little influenced by remaining uncertainties in $\alpha$.
We conclude that unobscured UV luminosity density (and hence unobscured star-formation density) 
peaks at $z \simeq 2.5 - 3$, when the Universe was $\simeq 2.5$\,Gyr old. Whether 
or not this coincides with the peak in {\it total} cosmic star-formation rate density ($\rho_{SFR}$) 
depends on the results of ongoing efforts to determine the level and evolution of 
dust obscuration at this epochs.

{}

\section*{Acknowledgments}
SP acknowledges the support of the University of Edinburgh via the Principal's Career Development Scholarship. 
JSD acknowledges the support of the European Research Council via the award of an Advanced Grant,  and the contribution of the EC FP7
SPACE project ASTRODEEP (Ref.No: 312725). RJM and AM 
acknowledge the support of the European Research Council via the award of a Consolidator Grant (PI McLure). 
This work is based in part on observations made with the NASA/ESA Hubble Space Telescope, which is operated by the 
Association of Universities for Research in Astronomy, Inc., under NASA contract NAS5-26555. 
This work is also based in part on observations made with the Spitzer Space Telescope, which is operated by the 
Jet Propulsion Laboratory, California Institute of Technology under NASA contract 1407. 
This work uses data taken with the Hawk-I instrument on the European Southern Observatory (ESO) Very Large Telescope from 
ESO programme: 092.A-0472. This work is based in part on data products from observations made with 
ESO Telescopes at the La Silla Paranal Observatories under ESO programme ID 179.A-2005 and on data products produced by TERAPIX and the 
Cambridge Astronomy survey Unit on behalf of the UltraVISTA consortium. This study was based in part 
on observations obtained with MegaPrime/MegaCam, a joint project of CFHT and CEA/DAPNIA, at the Canada-France-Hawaii
Telescope (CFHT) which is operated by the National Research Council (NRC) of Canada, the Institut National des
Science de l'Univers of the Centre National de la Recherche Scientifique (CNRS) of France, and the University of Hawaii.
This work is based in part on data products produced at TERAPIX and the Canadian Astronomy Data Centre as
part of the Canada-France-Hawaii Telescope Legacy Survey, a collaborative project of NRC and CNRS. 

\appendix

\section{Photometric redshift accuracy and reliability}

In this appendix we provide some additional details on the reliability and accuracy of 
our photometric redshifts in the three survey fields utilised in this study, and assess how 
our results compare with other recently published photometric redshift catalogues.

To estimate the accuracy of the photometric-redshift estimation procedure, we compare our photometric 
redshifts with their spectroscopic counterparts, for the subsamples of galaxies for which high-quality 
spectroscopic redshifts are known.

Following standard practice, we use the following statistics to quantify the 
accuracy and reliability of the photometric redshifts.

First, the basic scatter, $\sigma$, around the $z_{phot}:z_{spec}$ line is defined as: 

\begin{equation} 
\centering
\sigma=rms[\Delta{z}/(1+z_{spec})]
\end{equation}

\noindent
where $\Delta{z}=z_{phot}-z_{spec}$.

Second, `catastrophic outliers' are defined as galaxies for which:

\begin{equation} 
\centering
|\Delta{z}|/(1+z_{spec}) > 0.15.
\end{equation}

Third, the scatter can be recalculated after exclusion of the catastrophic outliers (in order to 
estimate the tightness of the core $z_{phot}:z_{spec}$ relation); this measure of scatter is usually denoted
as $\sigma_S$.

Fourth, an alternative measure of scatter, that minimizes the impact of (but does not require the removal of) 
catastrophic outliers, is the normalised median absolute deviation of $\Delta{z}$ defined as:

\begin{equation} 
\centering
\sigma_{NMAD}=1.48\times{median(\frac{|\Delta{z}|}{1+z_{spec}})}.
\end{equation}

As described in Section 3, we have assembled sub-samples of galaxies with high-quality spectroscopic 
redshifts within each of the three fields. In the HUDF field there are 218 such galaxies, and we obtained 
acceptable photometric redshifts (i.e. $\chi^2 < 50$) for 210 of these. The $z_{phot}:z_{spec}$ plot 
for this sub-sample of 210 galaxies is shown in the upper panel of Fig.\,A1, 
with the normalized redshift error, $\Delta z / (1 + z_{spec})$, plotted against $z_{spec}$ 
shown in the lower panel. The values of the four aforementioned statistics are given in the upper panel.
Analogous plots are then shown for the corresponding subsamples of 2677 galaxies in CANDELS/GOODS-S
(Fig.\,A2) and 1671 galaxies in UltraVISTA/COSMOS (Fig.\,A3). In all three fields the spectroscopic redhsifts 
provide coverage from $z \simeq 0$ to at least $z \simeq 5$, and the outlier/scatter statistics 
are consistent and competitive with the accuracy of the very best photometric redshifts as reported
elsewhere in the recent literature; outlier fraction is always significantly lower than $5$\%, and $\sigma_{NMAD} \simeq 0.026$
in all three fields.

\begin{figure}
\centering
\includegraphics[width=0.47\textwidth]{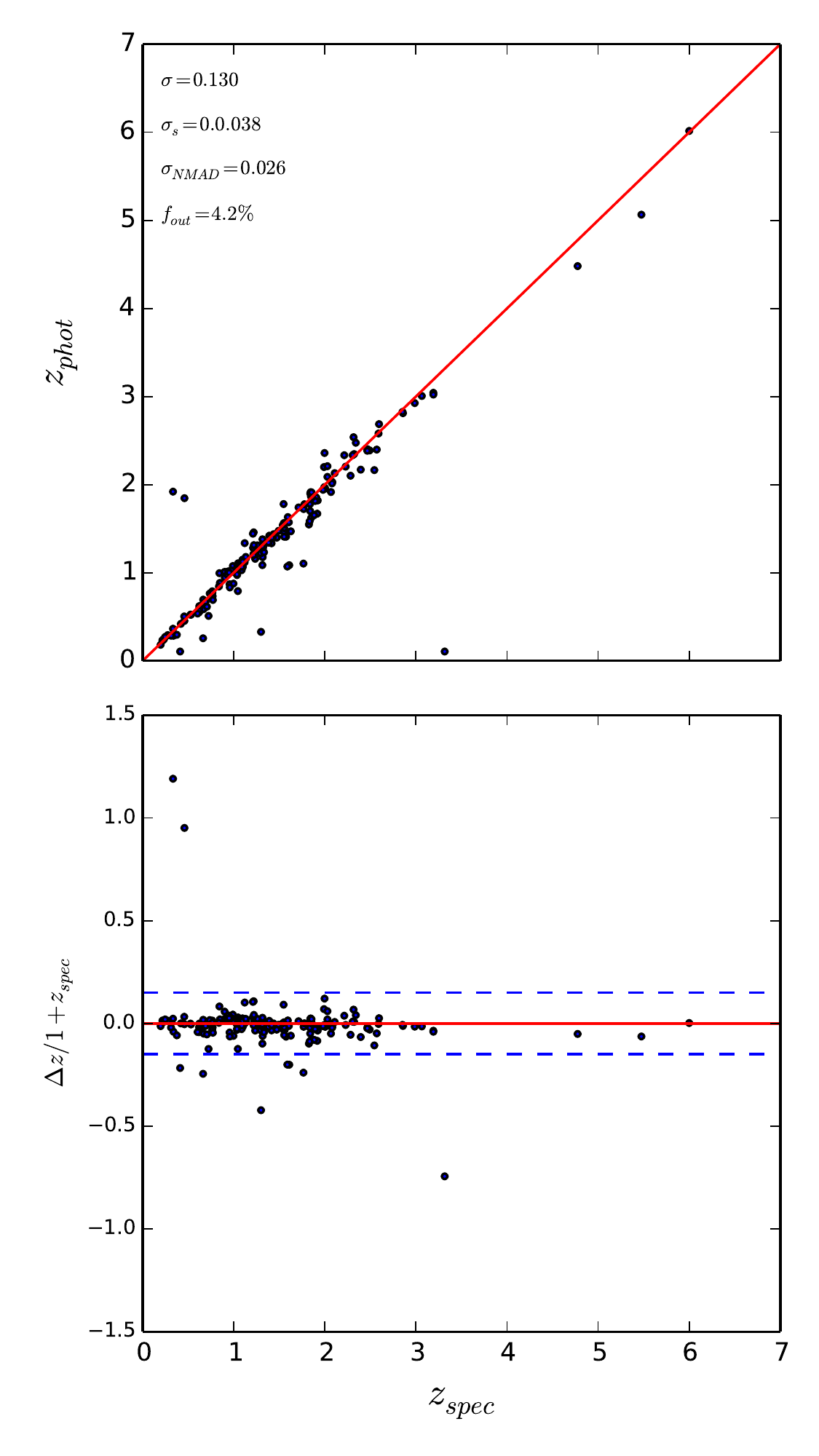}
\caption{$z_{phot}$ versus $z_{spec}$ for the 210 galaxies in the HUDF field 
with high-quality spectroscopic redshifts and acceptable photometric redshifts ($\chi^{2}<50$). The 
outlier/scatter measurements are given in the upper panel, while the lower panel 
redshift error as a function of $z_{spec}$, with the dashed blue lines 
inidcating the 0.15 boundary used to define the catastrophic outliers.}
\end{figure}

\begin{figure}
\centering
\includegraphics[width=0.47\textwidth]{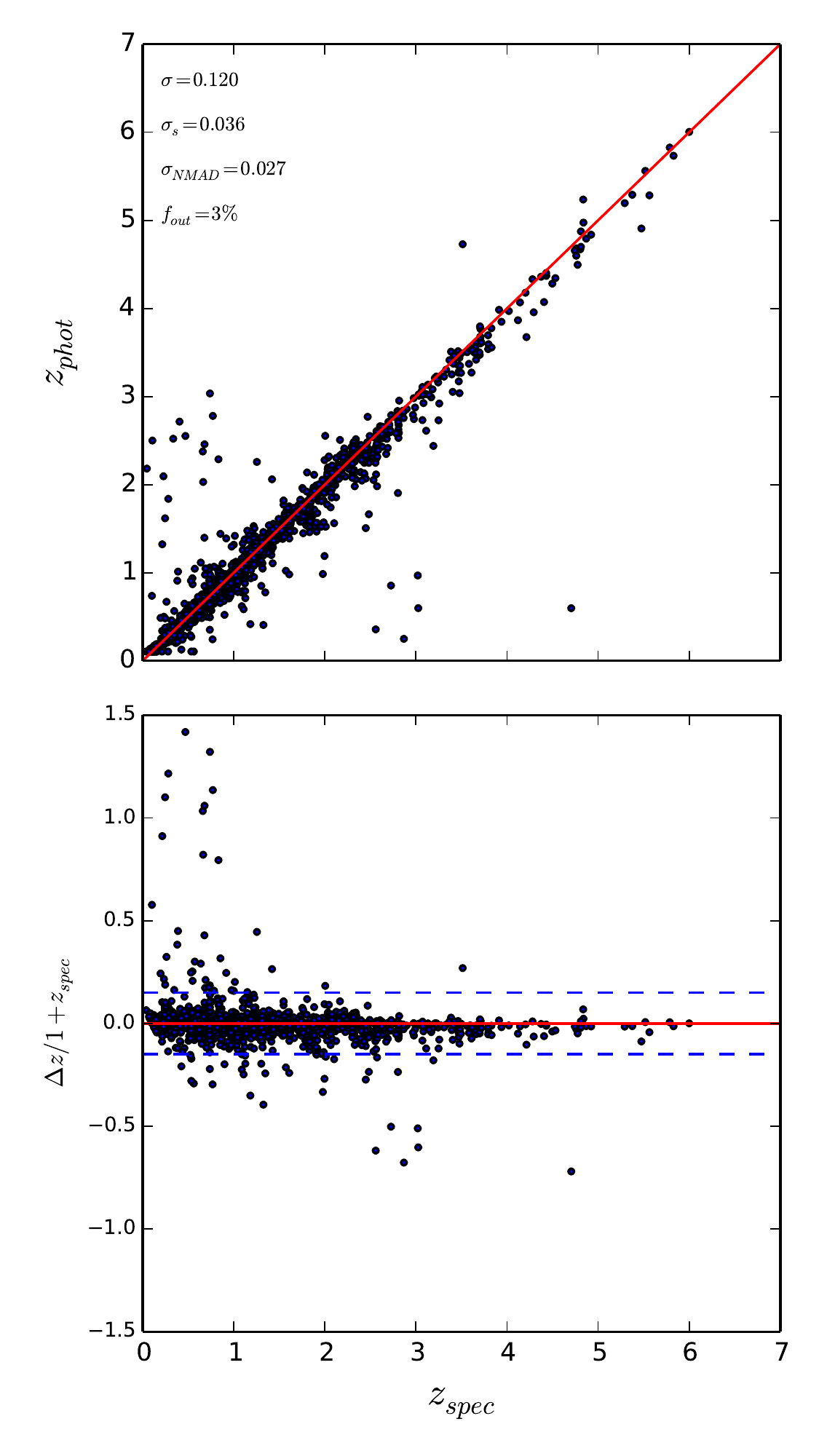}
\caption{$z_{phot}$ versus $z_{spec}$ for the 2677 galaxies in the CANDELS/GOODS-S field 
with high-quality spectroscopic redshifts and acceptable photometric redshifts ($\chi^{2}<50$). The 
outlier/scatter measurements are given in the upper panel, while the lower panel 
redshift error as a function of $z_{spec}$, with the dashed blue lines 
inidcating the 0.15 boundary used to define the catastrophic outliers.}
\end{figure}

\begin{figure}
\centering
\includegraphics[width=0.47\textwidth]{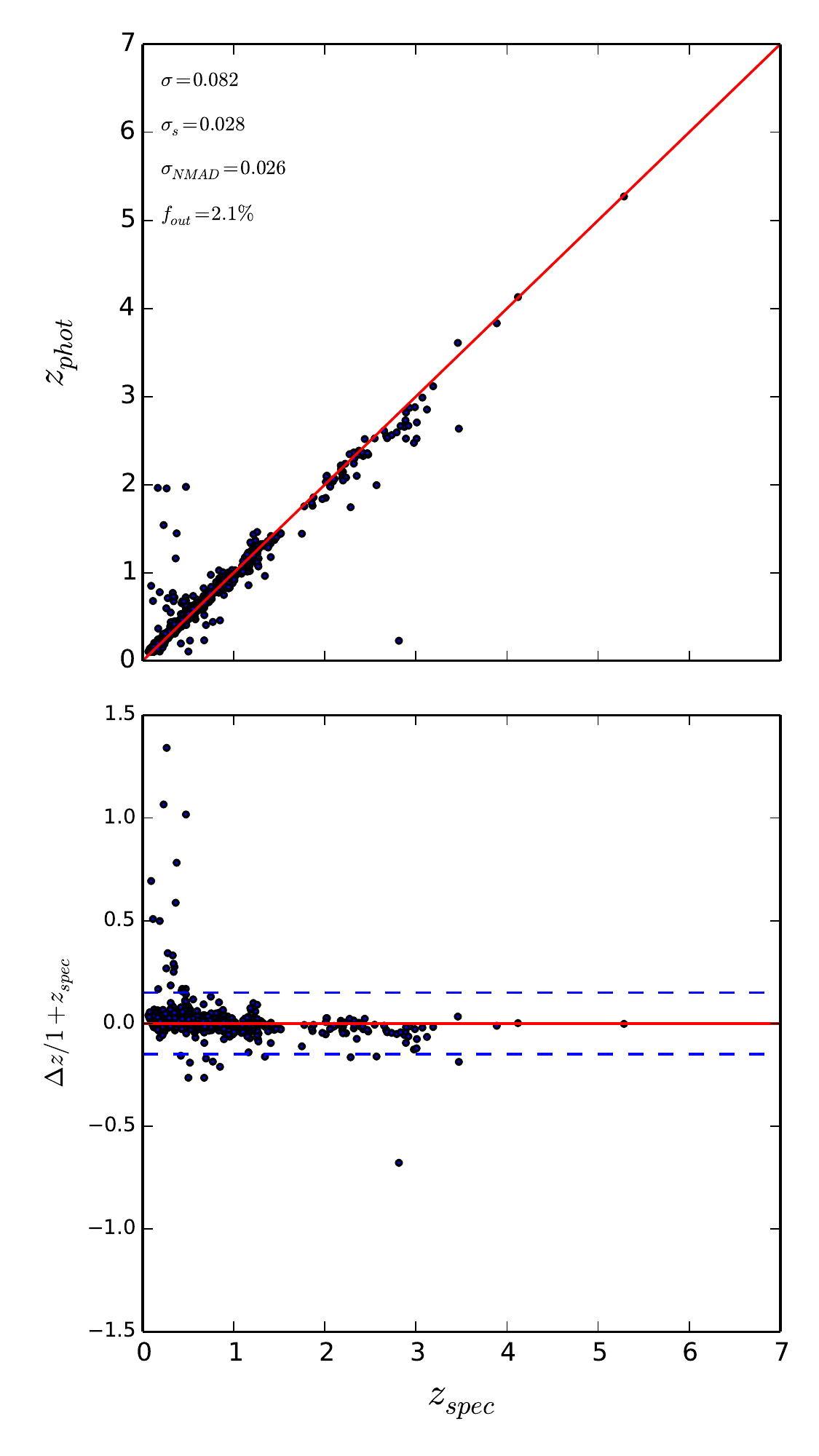}
\caption{$z_{phot}$ versus $z_{spec}$ for the 1671 galaxies in the UltraVISTA/COSMOS field 
with high-quality spectroscopic redshifts and acceptable photometric redshifts ($\chi^{2}<20$). The 
outlier/scatter measurements are given in the upper panel, while the lower panel 
redshift error as a function of $z_{spec}$, with the dashed blue lines 
inidcating the 0.15 boundary used to define the catastrophic outliers.}
\end{figure}

Within the CANDELS/GOODS-S field, an alternative set of photometric redshifts has recently been 
released by the 3D-HST team (Skelton et al. 2014). In Fig.\,A4 we plot our own $z_{phot}:z_{spec}$ results 
for this field (for the same 2677 galaxies shown in Fig.\,A2) along with the corresponding results 
as derived from the 3D-HST photometric redshift catalogue, and in Table\,A1 we compare the resulting 
outlier/scatter statistics. Clearly these two photometric redshift catalogues are of comparably 
high quality, although we note that the outlier fraction achieved here is significantly lower, 
possibly because the 3D-HST photometric catalogue does not contain the $HST$ $Y$-band imaging.
The slightly smaller $\sigma_{NMAD}$ achieved by the 3D-HST team appears to result from improved accuracy 
at low redshifts, possibly driven by their inclusion of medium-band ground-based Subaru imaging. However,
at the redshifts of interest in the present study ($z > 1.5$), our own measurements yield a slightly 
smaller scatter than is achieved by using the 3D-HST catalogue.

\begin{figure}
\centering
\includegraphics[width=0.47\textwidth]{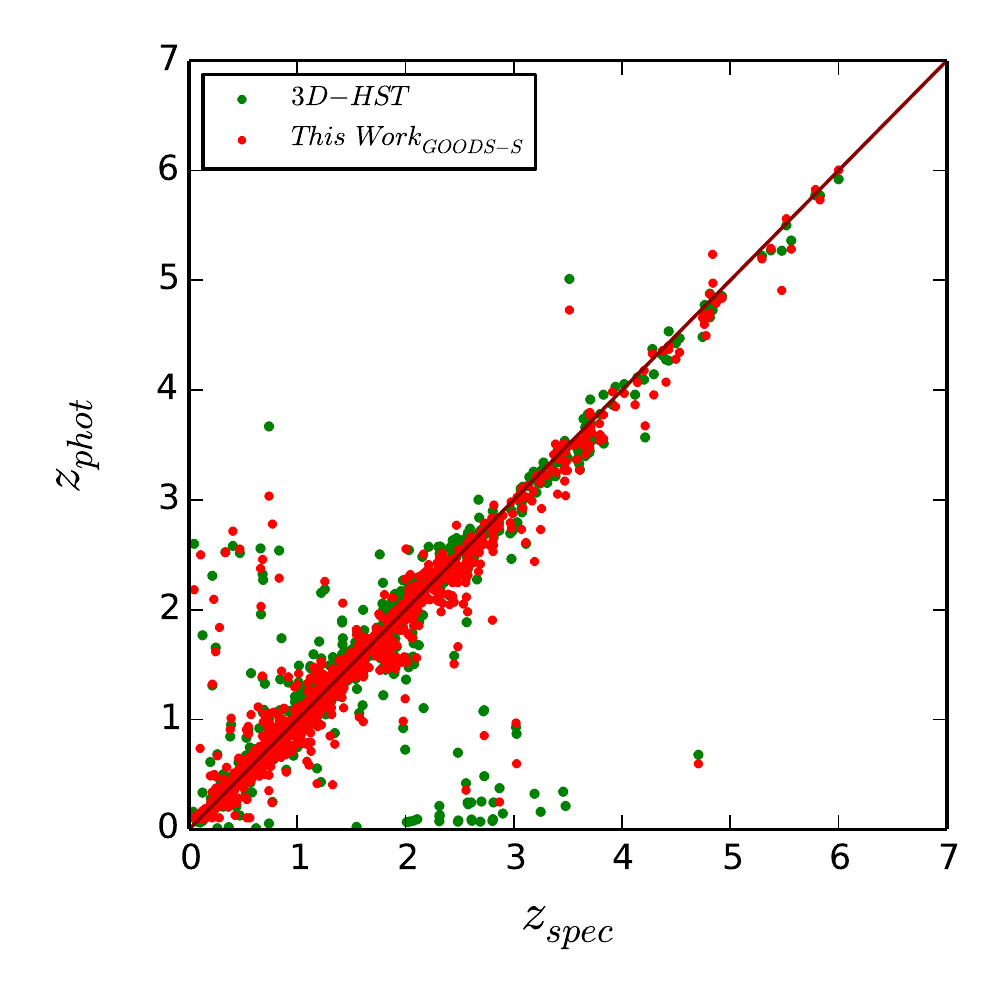}
\caption{A comparison of the reliability/accuracy of our photometric redshifts in the 
CANDELS/GOODS-S field with that achieved by the 3D-HST team (Skelton et al. 2014).
Our results (red points) are overlaid on the results derived from the public catalogue 
released by the 3D-HST team (green points) for the same spectroscopic sub-sample of 2677 
galaxies as previously discussed and presented in Section 3 and Fig.\,A2. Outlier fractions 
and scatter statistics are summarized in Table A1, and discussed in the text.}
\end{figure}

\begin{table} 
\centering
\caption{The reliability and accuracy of the photometric redshifts for 
galaxies in the CANDELS/GOODS-S field, as achieved here (see Section 3) and alternatively by utilising the 
the public 3D-HST photometric redshift catalogue (Skelton et al. 2014).
$\sigma$, $\sigma_{NMAD}$ and $f_{out}$ (see text) have been calculated for the high-quality spectroscopic 
sub-sample of 2677 galaxies (see Fig.\,A4). Our own results yield a lower 
outlier fraction, possibly because the 3D-HST photometric catalogue does not contain the $HST$ $Y$-band imaging.
The slightly smaller $\sigma_{NMAD}$ achieved by the 3D-HST team appears to result from improved accuracy 
at low redshifts, possibly driven by their inclusion of medium-band ground-based Subaru imaging. However,
at the redshifts of interest in the present study ($z > 1.5$), our own measurements yield a slightly 
smaller scatter than is achieved by using the 3D-HST catalogue.}
\begin{tabular}{l l l l l}
\hline
$Group$ &$\sigma$ & $\sigma_S$ & $\sigma_{NMAD}$ & $f_{out}$ \\
\hline 
\hline
This Work & 0.120 & 0.036 & 0.027 & 3.0\%\\ 
3D-HST & 0.135 & 0.027 & 0.013 & 4.3\% \\
\hline
\hline
\end{tabular}

\end{table}

Similarly, within the HUDF field, an alternative set of photometric redshifts has recently been 
released by Rafelski et al. (2015). In Fig.\,A5 we plot our own $z_{phot}:z_{spec}$ results 
for this field along with the corresponding results 
as derived from the Rafelski et al. (2015) photometric redshift catalogue. Here we are plotting results for 207 
galaxies (because 3 of the 210 galaxies plotted in Fig.\,A1 do not have photometric redshifts in the 
Rafelski et al. (2015) catalogue). We also include the results from both of the alternative redshift estimation 
techniques used by Rafelski et al. (2015), which are based on the Bayesian Photometric Redshift (BPZ; Benitez 2000)
and Easy and Accurate $z_{phot}$ from Yale (EAZY; Brammer, van Dokkum \& Coppi 2008) algorithms. 
The former uses a set of PEGASE SED models which have been re-calibrated based 
on observed photometry and spectroscopic redshifts from the FIREWORKS catalogue; emission lines are included and 
luminosity functions observed in COSMOS, GOODS-MUSIC and the UDF are used as priors. The latter method 
uses the default EAZY SEDs, with emission lines again included.
In Table\,A2 we again compare the resulting 
outlier/scatter statistics. Despite the fact that the Rafelski et al. (2015) results include new WFC3/UVIS 
photometry, again it can be seen that all three sets of results are competitive (presumably because our own calculations
utilise the VIMOS $U$-band ground-based photometry, minimizing the additional impact of the new UVIS data). 
Indeed, our outlier fraction is lower than yielded by the Rafelski et al. (2015) EAZY results which, as in our own calculations,
avoid the use of luminosity function priors, and our catalogue yields the lowest value of $\sigma_{NMAD}$.

\begin{figure}
\centering
\includegraphics[width=0.47\textwidth]{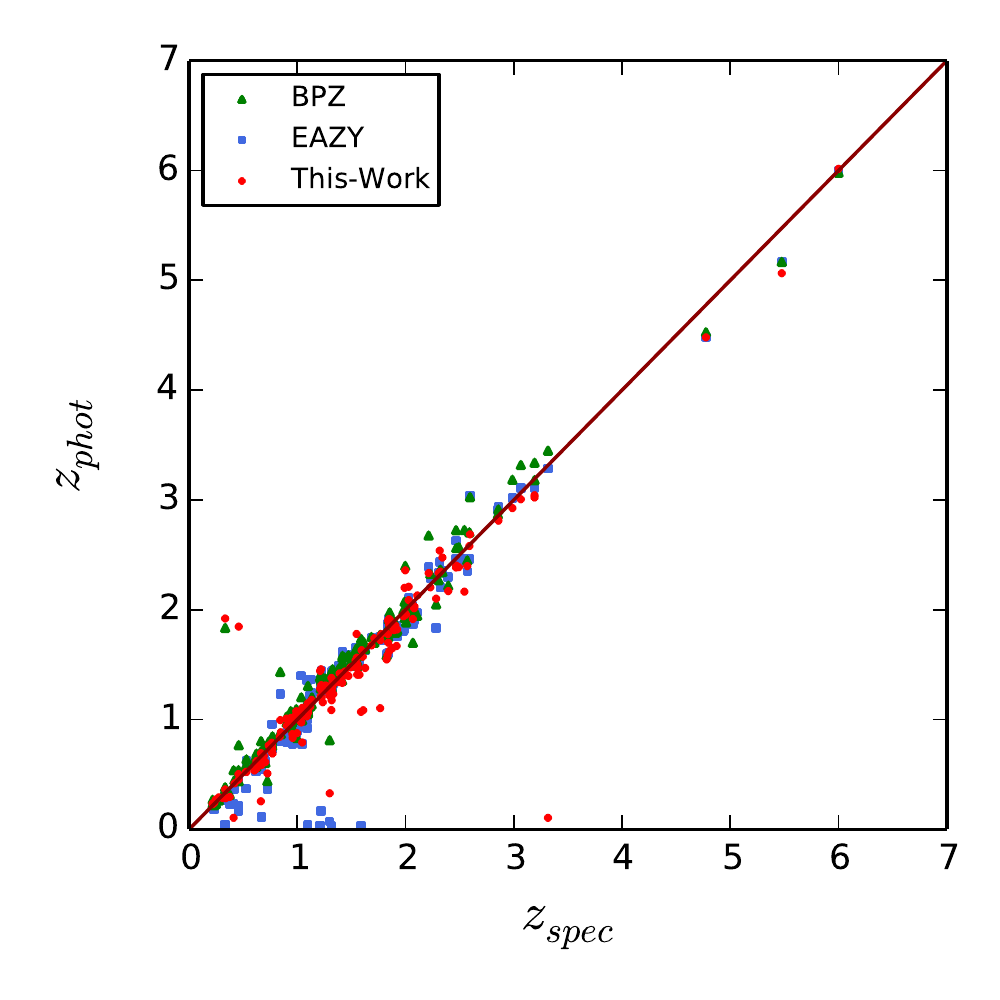}
\caption{A comparison of the reliability/accuracy of our photometric redshifts in the 
HUDF field with that achieved by the Rafelski et al. (2015).
Our results (red points) are overlaid on the results derived using the BPZ (green triangles) and EAZY 
(blue triangles) algorithms 
by Rafelski et al. (2015) for 207 of the 210 galaxies in the HUDF spectroscopic sub-sample 
as previously discussed and presented in Section 3 and 
plotted in Fig.\,A1. Outlier fractions 
and scatter statistics are summarized in Table A2, and discussed in the text.}
\end{figure}

\begin{table}
\centering
\caption{The reliability and accuracy of the photometric redshifts for 
galaxies in the HUDF, as achieved here (see Section 3) and alternatively by utilising the 
the new public photometric redshifts released by Rafelski et al. (2015).
$\sigma$, $\sigma_{NMAD}$ and $f_{out}$ (see text) have been calculated for 207 
of the galaxies with high-quality spectroscopic redshift in the HUDF (see Fig.\,A5). 
Our own results produce  a lower outlier fraction than yielded by the Rafelski et al. (2015) 
EAZY results which, as in our own calculations, avoid the use of luminosity function priors.
Moreover, our catalogue yields the lowest value of $\sigma_{NMAD}$.}
\begin{tabular}{l l l l l}
\hline
  $Algorithm$ & $\sigma$ &  $\sigma_S$ & $\sigma_{NMAD}$ & $f_{out}$ \\
\hline\hline
  BPZ & 0.093 & 0.038 & 0.033 & 2.4$\%$\\
  EAZY & 0.108 & 0.043 & 0.033 & 6.3$\%$\\
  This Work & 0.131 & 0.037 & 0.026 & 4.3$\%$\\
\hline\end{tabular}
\end{table}

\end{document}